\documentclass[a4paper,11pt]{article}

\usepackage{caption}
\usepackage{subcaption}
\usepackage{jinstpub} 
\usepackage{multirow}
\usepackage{lineno}

\title{\boldmath Batch VUV4 Characterization for the SBC-LAr10 scintillating bubble chamber}

\author[a,1]{H. Hawley-Herrera,\note{Corresponding author.}}
\author[b]{E.~Alfonso-Pita,}
\author[c]{E.~Behnke,}
\author[d]{M.~Bressler,}
\author[a]{B.~Broerman,}
\author[a]{K.~Clark,}
\author[a]{J.~Corbett,}
\author[e,f]{C.E.~Dahl,}
\author[a]{K.~Dering,}
\author[a]{A.~de St. Croix,}
\author[g]{D.~Durnford,}
\author[h]{P.~Giampa,}
\author[i]{J.~Hall,}
\author[j]{O.~Harris,}
\author[d]{N.~Lamb,}
\author[k]{M.~Laurin,}
\author[c]{I.~Levine,}
\author[l]{W.H.~Lippincott,}
\author[e]{X.~Liu,}
\author[a]{N.~Moss,}
\author[d]{R.~Neilson,}
\author[g]{M.-C.~Piro,}
\author[d]{D.~Pyda,}
\author[e]{Z.~Sheng,}
\author[a]{G.~Sweeney,}
\author[b]{E.~Vázquez-Jáuregui,}
\author[m]{S.~Westerdale,}
\author[l]{T.J.~Whitis,}
\author[a]{A.~Wright,}
\author[a]{E. Wyman,}
\author[l]{R.~Zhang}

\affiliation[a]{Queen`s University, Department of Physics, Engineering Physics, and Astronomy, Queen’s University, Kingston, K7L 3N6, Canada}
\affiliation[b]{Instituto de Física, Universidad Nacional Autónoma de México, A.P. 20-364, Ciudad de México 01000, México.}
\affiliation[c]{Department of Physics and Astronomy, Indiana University South Bend, South Bend, IN 46634, USA}
\affiliation[d]{Department of Physics, Drexel University, Philadelphia, PA 19104, USA}
\affiliation[e]{Department of Physics and Astronomy, Northwestern University, Evanston, IL 60208, USA}
\affiliation[f]{Fermi National Accelerator Laboratory, Batavia, IL 60510, USA}
\affiliation[g]{Department of Physics, University of Alberta, Edmonton, T6G 2E1, Canada}
\affiliation[h]{TRIUMF, Vancouver, BC V6T 2A3, Canada}
\affiliation[i]{SNOLAB, Lively, Ontario, P3Y 1N2, Canada}
\affiliation[j]{Northeastern Illinois University, Chicago, IL 60625, USA}
\affiliation[k]{Département de Physique, Université de Montréal, Montréal, H3T 1J4, Canada}
\affiliation[l]{Department of Physics, University of California Santa Barbara, Santa Barbara, CA 93106, USA}
\affiliation[m]{Department of Physics and Astronomy, University of California, Riverside, Riverside, CA 92507, USA}
\emailAdd{hawleyherrera.h@queensu.ca}

\abstract{The Scintillating Bubble Chamber (SBC) collaboration purchased 32 Hamamatsu VUV4 silicon photomultipliers (SiPMs) for use in SBC-LAr10, a bubble chamber containing 10~kg of liquid argon. A dark-count characterization technique, which avoids the use of a single-photon source, was used at two temperatures to measure the VUV4 SiPMs breakdown voltage ($V_{\text{BD}}$), the SiPM gain ($g_{\text{SiPM}}$), the rate of change of $g_{\text{SiPM}}$ with respect to voltage ($m$), the dark count rate (DCR), and the probability of a correlated avalanche (P$_{\text{CA}}$) as well as the temperature coefficients of these parameters. A Peltier-based chilled vacuum chamber was developed at Queen's University to cool down the Quads to $233.15\pm0.2$~K and $255.15\pm0.2$~K with average stability of $\pm20$~mK. An analysis framework was developed to estimate $V_{\text{BD}}$ to tens of mV precision and DCR close to Poissonian error. The temperature dependence of $V_{\text{BD}}$ was found to be $56\pm2$~mV~K$^{-1}$, and $m$ on average across all Quads was found to be $(459\pm3(\rm{stat.})\pm23(\rm{sys.}))\times 10^{3}~e^-$~PE$^{-1}$~V$^{-1}$. The average DCR temperature coefficient was estimated to be $0.099\pm0.008$~K$^{-1}$ corresponding to a reduction factor of 7 for every 20~K drop in temperature. The average temperature dependence of P$_{\text{CA}}$ was estimated to be $4000\pm1000$~ppm~K$^{-1}$. P$_{\text{CA}}$ estimated from the average across all SiPMs is a better estimator than the P$_{\text{CA}}$ calculated from individual SiPMs, for all of the other parameters, the opposite is true. All the estimated parameters were measured to the precision required for SBC-LAr10, and the Quads will be used in conditions to optimize the signal-to-noise ratio.}


\begin{document}
\maketitle
\flushbottom

\section{Introduction}
\label{sec:intro}
The goal of the Scintillating Bubble Chamber (SBC) collaboration is to detect GeV-scale dark matter and MeV-scale reactor antineutrinos. As a first step, the SBC collaboration is constructing a detector called SBC-LAr10 at Fermilab containing 10~kg of liquid argon (LAr) doped with xenon for engineering and calibration studies. The xenon doping shifts the LAr scintillation wavelength from 128~nm to 178~nm, characteristic of xenon scintillation \cite{neumeier2015}. Compared to Freon-based bubble chambers for dark matter searches like PICO \cite{amole2019}, the use of a scintillating target material improves background rejection. There is an intrinsic reduction in the probability of an electron-recoil-induced bubble, as demonstrated in ref.~\cite{baxter2017}. The collected scintillation light provides additional information about the energy deposited in the detector. For these detectors with low energy regions of interest, a signal is a bubble nucleation with no coincident scintillation photon detected.

The concept for a scintillating bubble chamber has been demonstrated in a 30-g liquid xenon (LXe) bubble chamber built at Northwestern University \cite{baxter2017}. In the LXe chamber, a single photomultiplier tube (PMT) was used for light collection. However, with the increased size and complexity of the full SBC-LAr10 detector, silicon photomultipliers (SiPM) were chosen as an alternative solution. SiPMs have several advantages over PMTs: they are more compact, are resistant to magnetic fields, perform at lower voltages, have no special requirements to retain their full functionality under cryogenic temperatures, and can recover without damage from over-exposure to light \cite{gundacker2020}. All of these characteristics make them a suitable light collection device for SBC. For more information about the other aspects of SBC, see \cite{alfonso-pita2022, alfonso-pita2023, hawley-herrera2024}.

SiPMs consist of an array of tens to hundreds of thousands of single photon avalanche diodes (SPADs) connected to a single common node via a quenching resistor. Applying a voltage to SPADs which is higher than the intrinsic breakdown voltage, $V_{\text{BD}}$, causes the SiPM to enter the avalanche regime. A successfully absorbed photon in a single SPAD operating in this regime will generate an electron-hole (e-h) pair that can start an avalanche of e-h pairs with a gain, $g_{\text{SiPM}}$, between $10^5-10^7~e^-$ \cite{acerbi2015,piemonte2016}. Once a SPAD is triggered or fired, it is unable to detect any further photons until it at least partially recharges. Multiple fired SPADs are required for multiple photo-electron (or PE) detection up to the number of SPADs in a SiPM. As a consequence, the SiPM can show non-linear artifacts as the shape of the pulse is dependent on the number of fired cells, and the emitted charge is not proportional to the number of captured photons if there are more photons than SPADs \cite{seifert2009}. Independently generated e-h pairs can be generated thermally inside a SPAD. The number of SPADs fired by thermal e-h pairs is Poissonian-distributed and characterized by the dark count rate (or DCR). DCR units are usually documented as Hz~mm$^{-2}$ to account for the effective area of a SiPM. Temperatures lower than the boiling point of liquid xenon (165~K), DCR dependence for temperature reduces significantly and direct band-to-band tunnelling takes over \cite{gola2019}.

A consequence of the solid-state nature of SiPMs is correlated avalanches (or CAs). CAs are created by cross-talk photons emitted due to the release of energy stored inside the SPAD after an avalanche is triggered. They are non-Poissonian in nature and can be characterized by their probabilities following an avalanche, $\text{P}_{\text{CA}}$, as a function of temperature and voltage \cite{rosado2015, du2008}. Cross-talk photons, depending on their final interaction point and physical origin, can be further categorized into different groups. When a cross-talk photon is detected instantaneously relative to the original avalanche, it is referred to as direct cross-talk (DiCT), and when the process is delayed, it is referred to as delayed cross-talk (DeCT) \cite{acerbi2015}. When a DeCT photon triggers the same SPAD before it fully recharges, an apparent fractional PE avalanche (relative to a fully charged SPAD) can be emitted. These pulses are referred to as optically-induced afterpulsing (AP). Throughout this paper, the concept of prompt correlated avalanches (PCA) is used instead of DiCT. DiCT is reserved for situations in which the knowledge of the origin of a CA is known, while PCA describes any CA that is simultaneous within the context of the experiment. Similarly, delayed correlated avalanche (or DCA) is used instead of DeCTs. For further readings about the physical mechanisms and characterization of DiCT and DeCT see refs.~\cite{boulay2023, acerbi2019, aalseth2017}. For other SiPM aspects not covered in this paper, such as their internal construction, photon-detection efficiency, and other applications of SiPMs, see refs.~\cite{gola2019, acerbi2019, bisogni2019, riu2012}.

SBC purchased 32 Hamamatsu VUV4 multi-pixel photon counters (MPPCs also known as SiPMs) \footnote{\url{https://hamamatsu.com}} with quartz windows in 2019 for use in SBC-LAr10, optimized for LAr and LXe scintillation collection \cite{hoenk1992, acerbi2019}. Each Hamamatsu VUV4, or ``Quad'', contains four 5.95$\times$5.85~mm SiPMs with 13,923 SPADs each and a 60\% geometrical fill factor. For the SBC-LAr10 detector, these four SiPMs are wired together in parallel which reduces the required power supplies and amplifiers for each Quad.

The light collection system in SBC-LAr10 is used primarily as a veto against high-energy backgrounds for the dark matter search, and it does not require precise or high scintillation collection yields. The scintillation photons must travel through several optical interfaces (LAr to quartz, then quartz to hydraulic fluid, and, finally, to the SiPM), yielding an expectation of 2\% light collection efficiency. The exact veto scheme (e.g. minimum coincidence requirement, coincidence window, and minimum number of PE detected) will be informed by the scintillation backgrounds observed in-situ. These corrections should be well enough understood that systematic uncertainties in simulated spectra due to SiPM response are subdominant to uncertainties stemming from low-photon statistics and non-uniform light collection.

For SBC, the most important parameters for each Quad are the $V_{\text{BD}}$, the rate of change of $g_{\text{SiPM}}$ with respect to voltage (or $m$), and the average arrival time of the DCAs ($\bar{\tau}_{\text{DCA}}$). The first two are required to set the SiPMs at the optimal voltage for light collection and signal-to-noise ratio. $\bar{\tau}_{\text{DCA}}$ is required to set a limit on the time window needed before a trigger to avoid triggers caused by CAs. Additionally, the temperature dependence of the parameters needs to be well known to predict them at the range of expected operational temperatures. These parameters can be measured at two or more temperatures to characterize their temperature coefficients assuming linearity.

$V_{\text{BD}}$ can be estimated as the voltage intercept of the $g_{\text{SiPM}}$ vs.~voltage, $g_{\text{SiPM}} = m(V - V_{\text{BD}})$, or as the start of the avalanche region of the SiPM current as a function of voltage (I-V curve) \cite{nagai2019, dinu2017, acerbi2017}. These two methods complement each other, but combining both measurements in a single experimental setup presents challenges, particularly when the desired precision is below 1\%. The timing parameters can only be measured directly with pulse information, which requires the $g_{\text{SiPM}}$ method. For this reason, it was decided that the experimental setup focused on measuring dark count pulses, using the I-V information only for diagnostics. The benefit of using dark count pulses is no single-photon source is required, but this limits the characterization to temperatures where the DCR is not negligible.

Although only $V_{\text{BD}}$, $m$, and $\bar{\tau}_{\text{DCA}}$ are of particular interest for SBC-LAr10, the data gathered could be used to study the batch-to-batch variations in SiPMs provided by the manufacturer. Understanding these variations in Hamamatsu VUV4 with respect to the reported values would be helpful in the design of future experiments. To achieve this goal, each SiPM in a Quad is measured regardless of the SBC-LAr10 implementation, and a precision <0.5\% is required for statistically significant observation of the manufacturer spread. Doing so should not affect their use in SBC-LAr10 as all parameters can be estimated from individual SiPMs parameters. For example, the Quad $V_{\text{BD}}$ is equal to the smallest $V_{\text{BD}}$ of the individual Quads in the SiPM. The estimation and spread of DCR and P$_{\text{CA}}$ will also be included. 

During a preliminary quality control verification of the Quads, two Quads were found to be abnormal, and they were replaced by Hamamatsu two years after the acquisition date. These two new Quads throughout the test are referred to as ``batch two'' Quads while the other Quads are referred to as ``batch one''. This distinction provides an opportunity to test if there are significant differences between the manufacturer batches.

\section{Experimental Setup}
\label{sec:exp_setup}
\begin{figure}[htbp]
    \centering
    \includegraphics[width=0.55\textwidth]{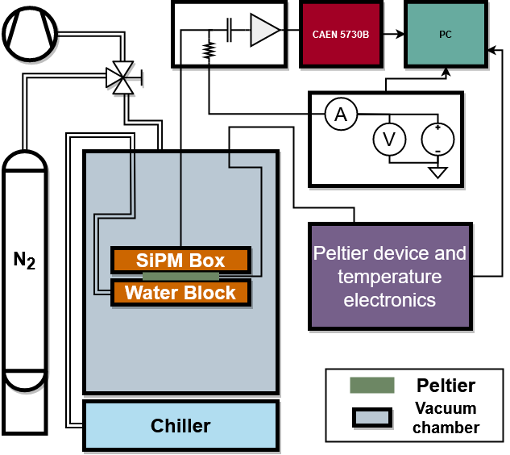}
    \caption{A simplified schematic of the experimental setup used to characterize the 32 SBC Hamamatsu VUV4 Quads. The temperature control system consists of the water block, the chiller, the Peltier device, and temperature electronics. The readout electronics comprise the custom-made TRIUMF amplifier, a CAEN 5730B digitizer, a Keithley 2000 multimeter, and a Keithley 6487 picoammeter shown by their respective symbols. The N$_2$ tank fills the chamber at the end of a measurement procedure to minimize damage to the SiPMs from water condensation.}
    \label{fig:sipm_cooler}
\end{figure}

The objective in the design of the characterization setup is to identify and minimize any sources of systematic error that might be misinterpreted as the underlying manufacturer distribution of the parameters. The most significant contributors identified are temperature variation and the readout electronics. Another less apparent contributor to errors is the measurement procedure. Improving the handling quality reduces mistakes and potential damage to the SiPMs which could be reflected in the parameters. This section is divided into three parts describing the justification for all the decisions made to standardize the system. The first one is dedicated to the temperature control system, along with the solutions to minimize temperature interface error. The second part is committed to the readout electronics consisting of the amplifier, the digitizer, and the voltage and current acquisition system. Finally, the measurement routine is summarized which includes the handling considerations and the acquisition measurement routine. A simplified schematic of the experimental setup can be seen in figure~\ref{fig:sipm_cooler}.

\subsection{Temperature system}
\label{sec:exp_temp}
The $V_{\text{BD}}$ and the DCR have been shown to be dependent on temperature \cite{gola2019, gallina2019}. Variations in the temperature measurement will be reflected in the uncertainty of the characterization, specifically on the DCR. The objective of the temperature control system is to reduce the temperature variation to levels where it becomes an insignificant contributor to the error of the parameters while also providing the ability to characterize the SiPMs at two different temperatures.
%A Peltier-based temperature control system was deemed as the only one capable of reaching low enough temperatures that still have reasonable dark noise rates while having low-temperature fluctuations and programmable temperatures with the conjunction of a PID controller. 

A SP2402-01AB Peltier device\footnote{\url{https://ii-vi.com/thermoelectrics/}} device was chosen to cool down a $48 \times 80 \times 23$~mm$^3$ copper SiPM holder box designed to hold two Quads at a time. Two NB-PTCO-160 B-class 4-wire PT100 resistive temperature devices\footnote{\url{https://www.te.com}} are inserted inside holes behind the Quads mounting positions in the box. Thermal contact between the box and the Quads is achieved with a folded indium foil. Thermal adhesives, greases, or sponges were not used as metallic dust or any residue could damage the Quads and potentially be dissolved in the liquid carbon tetrafluoride environment the SiPMs are later surrounded by when operating in SBC-LAr10 which would introduce undesired radioactive particle backgrounds. To further aid in the thermal contact between the Quads and the copper box, the Quads sit in a PCB with solid thermal contact with the box through coaxial connectors. Additionally, the clearance between the PCB and the box allowed the Quad to put additional pressure on the indium foil, further improving the thermal contact. 

A copper water block, directly attached to the Peltier device, is connected to a chiller using a water/glycol mixture (75\% water by volume) as the heat exchange medium to remove waste heat. The Peltier driver is a constant current source using a buck-driver configuration with a 6~A max current load and no measurable electrical interference in the SiPM readout electronics. The PT100 readout consists of three modified Adafruit MAX31865\footnote{\url{https://www.adafruit.com}} based PT100 RTD temperature sensor amplifiers where the reference resistor was replaced with a 350~$\Omega$ resistor with a $\pm 0.01\%$ accuracy rating and a temperature coefficient equal to 0.2~ppm$~^\circ$C$^{-1}$. A Teensy 4.1 microcontroller\footnote{\url{https://www.pjrc.com/teensy/}} interfaces the MAX31865 boards with the computer, and it runs the PID controller logic. 

\begin{figure}
    \centering
    \includegraphics[width=\textwidth]{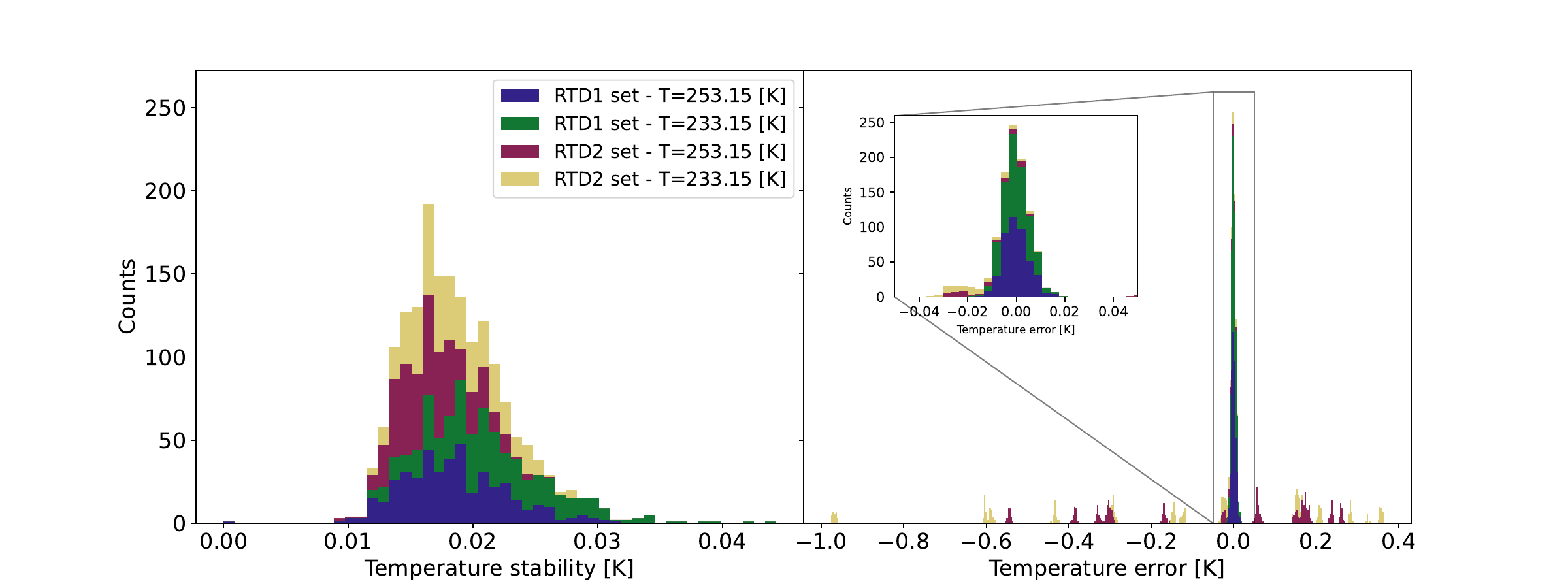}
    \caption{Right: the difference of the expected temperature (the PID temperature setpoint) with respect to the average of the measured temperature while data is taken across all Quads. RTD2 differed by up to 1~K from measurement to measurement because of a poor thermo-mechanical connection to the SiPM box. RTD1 is used as the final reporting temperature measurement device for both tested Quads. Left: the stability at 1 std of the temperatures while data taking.}
    \label{fig:readout:temps}
\end{figure}

The PT100s are placed inside the SiPM box using Kryonaut Extreme thermal grease to hold them in place and maximize thermal contact. Although two PT100s were used to measure temperature, only one was used as the PID control transducer and the final reporting temperature measuring device. The most stable PT100, RTD1, was chosen during the testing campaign, while RTD2 was kept for diagnostics. RTD2 temperature readings initially agreed with RTD1, but the installation procedure, later described, caused it to slide partially outside the box. Therefore, RTD2 no longer maintained a reasonable thermo-mechanical connection to the SiPM box, which was measured as a larger spread in its temperature measurements (see right-side of figure~\ref{fig:readout:temps}).  Keeping the PID device to a single PT100 guarantees that any temperature gradient between the Quad and the PT100 would remain consistent between measurements. 

A temperature calibration was never performed, with results relying on manufacturer uncertainties for the final reported temperatures. The temperature errors were calculated using NIST technical note 1297 as the base to get the equivalence of all the sources of error from the manufacturers to one standard deviation \cite{taylor1994}. The ADC counts to resistance were evaluated from the equations found in the MAX31865 technical document, and the resistance-to-temperature conversion was done using the ``T$<0^\circ$C ITS-90'' polynomial \cite{preston-thomas1990}. The resulting systematic uncertainty for the temperatures of interest is $\pm0.2$~K. A calibration procedure could have been employed to reduce the uncertainty down to the system temperature stability, see the left side of figure~\ref{fig:readout:temps}; however, anything more precise would only significantly decrease the DCR systematic uncertainty discussed in section~\ref{sec:results}.
% Reported temperatures can be done using the $V_{\text{BD}}$ and DCR. If their reported spread matches the expected spread from the temperature, it can conclusively test if the temperature variations have been under-reported. This test is discussed during section~\ref{sec:results}.

Measuring two Quads at the same time decreases the length of time required for the measurement campaign, but two Quads facing each other introduces an additional source of noise in the form of external cross-talk. External cross-talk is a source of CA caused by the emission of photons from another SiPM detected in the test SiPM \cite{boulay2023}, which will impact the measurements as a fictitious increase in the DCR and additional CAs. A 14-mm-thick block of HDPE was installed between the Quads to block the external cross-talk photons.

The difference between expected temperatures with respect to the average of the measured temperatures while taking data across all SiPMs can be seen on the right side of figure~\ref{fig:readout:temps} for both PT100s.

\subsection{Readout electronics}
\label{sec:readout_elecs}
% The readout electronics consist of a TRIUMF high bandwidth SiPM amplifier with an expected gain of 135 and an RMS noise figure of approximately equal to 700~$\mu$V. The SiPM waveforms were digitzed using a CAEN 5730B digitizer. While the SiPM voltage supply and SiPM current measurements were done with a calibrated Keithley 6487 picoammeter. Then, the SiPM applied voltages were measured using a calibrated Keithley 2000. Finally, the acquisition software also plays an additional role in the collection and rejection of the waveforms. The schematic of the connections can be seen in figure \ref{fig:readout_electronics}. This part of the text will discuss each of the components of the readout electronics, and logic behind the design decisions of the readout electronics such as the thresholds, and precision levels.
The readout electronics were chosen to keep the complexity of the measurement procedure to a minimum while maximizing accuracy and precision. The readout electronics had to be kept static across all different configurations. For example, the digitizer threshold, number of collected waveforms, and voltage ranges were kept constant across all Quads, testing temperatures, and voltages. To find the best set of parameters, a set of priors for the $V_{\text{BD}}$, DCR, and P$_{\text{CA}}$ found in ref.~\cite{gallina2019} were used to estimate the number of waveforms required to collect enough statistics for a maximum 0.5\% accuracy rating across all parameters. For further information, three additional testing Quads not part of the SBC Quads were used to construct and test a proto-version of the measurement procedure. The readout electronics include a single-channel custom-made TRIUMF amplifier, a CAEN 5730B digitizer, a Keithley 6487 serving as a voltage supply and the picoammeter, a Keithley 2000 multimeter, and the waveform pre-processing software. 

The single-channel custom-made TRIUMF SiPM amplifier is required to buffer and amplify the SiPM pulses to levels the digitizer can measure with a positive signal-to-noise ratio across all expected over-voltages. In addition to the amplifier stage, the TRIUMF amplifier internally distributes and conditions the voltage to the SiPM. The internal conditioning is achieved using several RC filters with a total equivalent resistance expected to be $R_{A} = 30600 \pm 60$~$\Omega$. After the conditioning stage, the voltage is distributed to the SiPM and the amplifier through a DC-blocking capacitor. Although the RC filtering circuit reduces electrical noise from the power supply, it introduces two additional sources of error, RC rise-time error and a voltage offset that follows the relationship: 
\begin{equation}
V_{\text{SiPM}}=V_{i} - I_{\text{SiPM}}R_{A},
\label{eq:sipm_voltage}
\end{equation}
where $V_{\text{SiPM}}$ is the voltage difference across the SiPM, $V_i$ is the applied voltage, and $I_{\text{SiPM}}$ is the measured SiPM current. The RC rise time error is reduced to insignificant levels if there is enough waiting time between changing the voltage and starting a measurement. For this amplifier with a SiPM, 90 seconds was measured enough for both voltage and currents to stabilize. The voltage offset correction was applied, as all parameters and variables were known at the time of the measurements.

A calibrated Keithley~6487\footnote{\url{https://www.tek.com/en/products/keithley/low-level-sensitive-and-specialty-instruments/series-6400-picoammeters}} picoammeter is used to supply $V_{\text{SiPM}}$ and read $I_{\text{SiPM}}$. The Keithley~6487 output voltage and current ranges were chosen based on the most significant contribution to the error budget for $V_{\text{SiPM}}$. From eq.~\ref{eq:sipm_voltage}, there are two sources of error. One source of error is the $V_i$ accuracy. The applied voltages are never expected to exceed 60~V at any point, as verified using the testing Quads. For the 100~V range, the maximum error is found at 60~V and is approximately equal to $\pm$2~mV at one standard deviation (std). The next contribution to the error budget for $V_{\text{SiPM}}$ is from the measured $I_{\text{SiPM}}$ accuracy rating. At the 20~$\upmu$A range, the maximum accuracy is $\sim$10~nA, and when an error is propagated with eq.~\ref{eq:sipm_voltage}, is equal to $\sim$1~mV below the multimeter accuracy.

An important consequence of identifying the main source of error for $V_{\text{SiPM}}$ is that any measured currents below 50~nA insignificantly increases $V_{\text{SiPM}}$ compared to the error of $\pm$2~mV. Therefore, the current measurement design can utilize conventional coaxial cables, and current offsets from parasitic resistance offsets are negligible \cite{keithleyinstruments1998}.

The next component in the electronic chain is the CAEN 5730B digitizer. This 14-bit, 8-ch digitizer has a 250 MHz bandwidth and 500 MS/s sampling rate with 0.5 V and 2 V selectable input ranges. The main parameter that had to be considered for the digitizer was the trigger threshold, which was decided to be several standard deviations away from the amplifier noise while keeping 100\% efficiency at acquiring single PE (SPE) waveforms across all temperatures and voltages of interest. The record length was set to be equal to 10~$\upmu$s which is double the size of the SBC-LAr10 region of interest. Finally, the voltage dynamic range was set to 0.5~V which was deemed wide enough to cover pulses with several degrees of CA. Otherwise, as described in appendix~\ref{app:mc_sipm}, the digitizer saturation can impact the accuracy of the timing parameters.

The $g_{\text{SiPM}}$ is 
\begin{equation}
    g_{\text{SiPM}} = \frac{\alpha}{q_E} \sum_{i=0}^N C[i] = \frac{\alpha}{q_E} A,
    \label{eq:sipm_gain}
\end{equation}
where $\alpha$ is the readout electronics gain, $q_E$ is the electron charge, and $C[i]$ are the digitizer counts. Under normal circumstances, eq.~\ref{eq:sipm_gain} is only required to calculate the gain while $\alpha$ can be estimated from the parameters of the digitizer, the gain of the amplifier, and the load resistance. During the characterization run, the TRIUMF amplifier malfunctioned and had to be repaired. The consequence of the repair is that $\alpha$ is not the same across all the SiPMs. The change in $\alpha$ was estimated using the ratio of the SPE gain (more on how the SPE gain was extracted in the section~\ref{sec:analysis}) from before and after the fix to the TRIUMF amplifier using SiPMs that were measured with both amplifiers. The ratio was estimated to be $0.955 \pm 0.003$.

% This paragraph might be removed if we do not find a way to measure it. An artificially-high 5% error was selected  to be higher than the expected
The original $\alpha$ was never measured due to the lack of specialized instrumentation to characterize it. An artificially high 5\% error was selected to be higher than the expected error from the manufacturer's uncertainties to account for temperature coefficients and possible batch-to-batch variations.

At the end of the acquisition chain, the interface software selects which waveforms should be saved to disk. The final selection criterion is that waveforms must have no previous corresponding waveform for at least 10~$\upmu$s. Without this condition, the first triggering pulse in a waveform might not be an independent pulse (see section~\ref{sec:analysis}). The length of the no-pulse criterion was chosen to maximize the efficiency of waveforms saved to disk across all temperatures and voltages. Longer times for the no-pulse criterion further guarantee that the sampled waveforms are independent at the cost of longer acquisition times.

\subsection{Measurement procedure}
\label{sec:exp_procedure}
One main objective of the measurement procedure is to reduce handling and exposure damage to the SiPMs. This includes electrostatic discharge (ESD) and air exposure degradation from humidity or oxygen. To maximize cleanliness and stability, gloves, masks, and ESD-safe materials were used throughout the procedure.

Exposure to air, or more specifically, humidity and dust, is expected to degrade the SiPM parameters. To minimize exposure, the SBC Quads were stored in the Hamamatsu-provided storage container and double-bagged with a desiccant packet. To further protect them, the double-bagged container was stored in a dry storage box which had the desiccant replaced whenever any visible colour change was observed. Finally, whenever the Quads were not used for an extended time, the bags would be thermally sealed.

\begin{figure}[htbp]
    \centering
    \begin{subfigure}[b]{0.45\textwidth}
        \centering
        \includegraphics[width=\textwidth, clip, trim={0 15cm 15cm 0}]{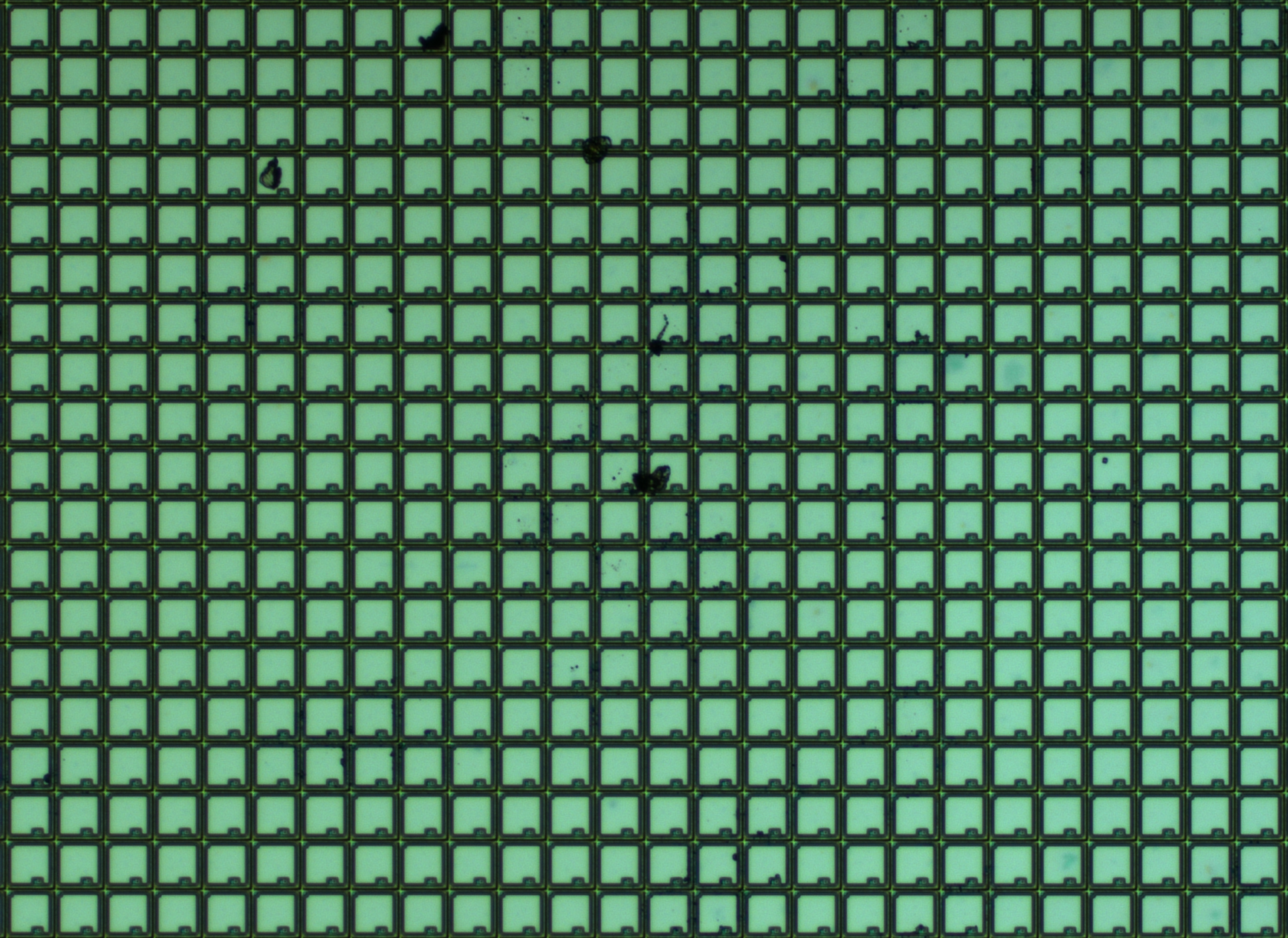}
        \caption{}
    \end{subfigure}
    \begin{subfigure}[b]{0.45\textwidth}
        \centering
        \includegraphics[width=\textwidth,trim=0 16cm 15cm 0,clip]{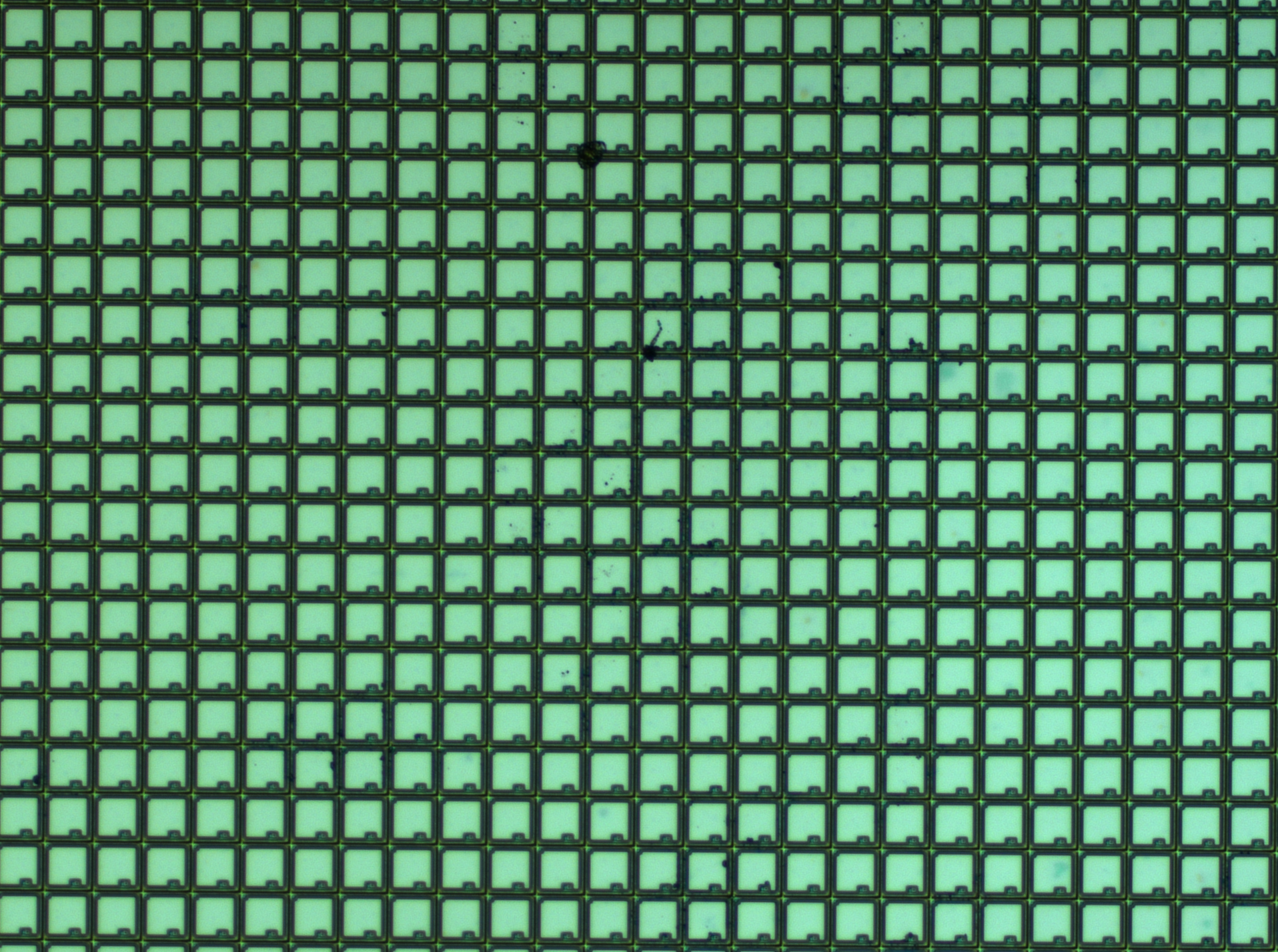}
        \caption{}
    \end{subfigure}
    \caption{(a) Pre-cleaned SiPM showing specks of dust covering some SPADs. (b) SiPM that has been cleaned with pressurized N$_2$. The leftover stains are suspected to be static-bound dust or physical damage. Photos were taken with a Nikon ECLIPSE Ni microscope.}
    \label{fig:cleaning}
\end{figure}

Before installing the Quads inside the SiPM box, they were cleaned using filtered high-purity pressurized nitrogen to remove any dust that might have accumulated on the surface of the quartz windows and the SiPMs. The Quads were held in a fixture that grounded all their pins to avoid any discharge that could harm their internal structure. The effectiveness of the cleaning procedure was verified with a test Quad, shown in figure~\ref{fig:cleaning}.

After cleaning the Quads, they were installed inside the SiPM testing setup. Immediately following the installation, the vacuum chamber was closed, and the vacuum pump was started. Once the pressure was reduced to the millibar range, the chiller was activated, taking roughly half an hour to stabilize the temperature. Once the temperature was stable, the Peltier device control was activated with the first temperature equal to 253.15~K. The first temperature setpoint took approximately an hour to stabilize. To continue the procedure, an indicator notified the user when the mean of the setpoint errors was in a $\pm10$~mK window from $\Delta T=0$~K.

Once a stable temperature was reached, the pulse acquisition could be started manually. The pulse acquisition routine consisted of two subroutines: the $V_{\text{BD}}$ estimation and the over-voltage (OV) acquisition. The $V_{\text{BD}}$ estimation subroutine gathers 20,000 waveforms at three different voltages: 53~V, 54~V, and 55~V. For each voltage, it acquired the pulses only after waiting 90~s after setting the voltage. Next, a first pass of the SPE estimation routine (more on the SPE routine in section~\ref{sec:analysis}) is used to estimate $V_{\text{BD}}$. Once the $V_{\text{BD}}$ was estimated, the OV acquisition routine acquires 200,000 waveforms at voltages that start at 2~V and end in 8~V in steps of 1~V above $V_{\text{BD}}$. As will be shown in section~\ref{sec:results}, the resulting voltages have a larger-than-expected dispersion. The original intention of the $V_{\text{BD}}$ subroutine was to standardize all the measurements to the same offset from $V_{\text{BD}}$, aiding any further analysis. It was later found that some assumptions introduced bias in the calculations, which is reflected in the dispersion in the voltage axis. The final analysis results should not be affected by the increased dispersion.

Once the OV subroutine was finished, the pulse acquisition routine was repeated for each SiPM in each Quad. After both Quads had their data collected, the temperature setpoint was changed to 233.15~K, and the pulse acquisition routine was started again. The entire pulse acquisition routine for both Quads at both temperatures took roughly 8 hours.

After the data collection, the Peltier is turned off, and the warming-up stage commences. No external heaters were required, as once the Peltier was off, heat would freely transfer from the water block to the SiPM box. The chiller temperature was also raised slightly to aid in the warming process. This stage lasts approximately 30 minutes before reaching a temperature above the dew point in the lab. Nitrogen is used to restore pressure inside the chamber to minimize any humidity ingress that could potentially condense on the Quads. 
% The pulse acquisition measurements were completed in 30 working days. The only major delay caused by the SiPM amplifier stage inadvertently stopped working half-way the second to last batch measurement. The replacement parts arrived a week later and the second to last batch had to be re-measured.

\section{Analysis}
\label{sec:analysis}
The analysis method plays a significant role in the characterization of the SiPMs. Biases and uncertainties must be minimized to prevent them from being mistaken for the underlying manufacturer variation. Additionally, the analysis must be relatively efficient, as the amount of data to process is in the order of terabytes. 
%The analysis is divided into five steps: waveform preprocessing, the single-photoelectron (SPE) extraction, the $V_{\text{BD}}$ estimation, the wave unfolding, and the timing distribution estimation. To aid in the analysis description through the paper, a single SiPM (Quad with ID 200 and SiPM 1) is used throughout as an example unless otherwise noted. This analysis was also tested against a Monte Carlo SiPM which is presented in appendix~\ref{app:mc_sipm}. 

\begin{figure}[htbp]
    \centering
    \includegraphics[width=0.55\textwidth]{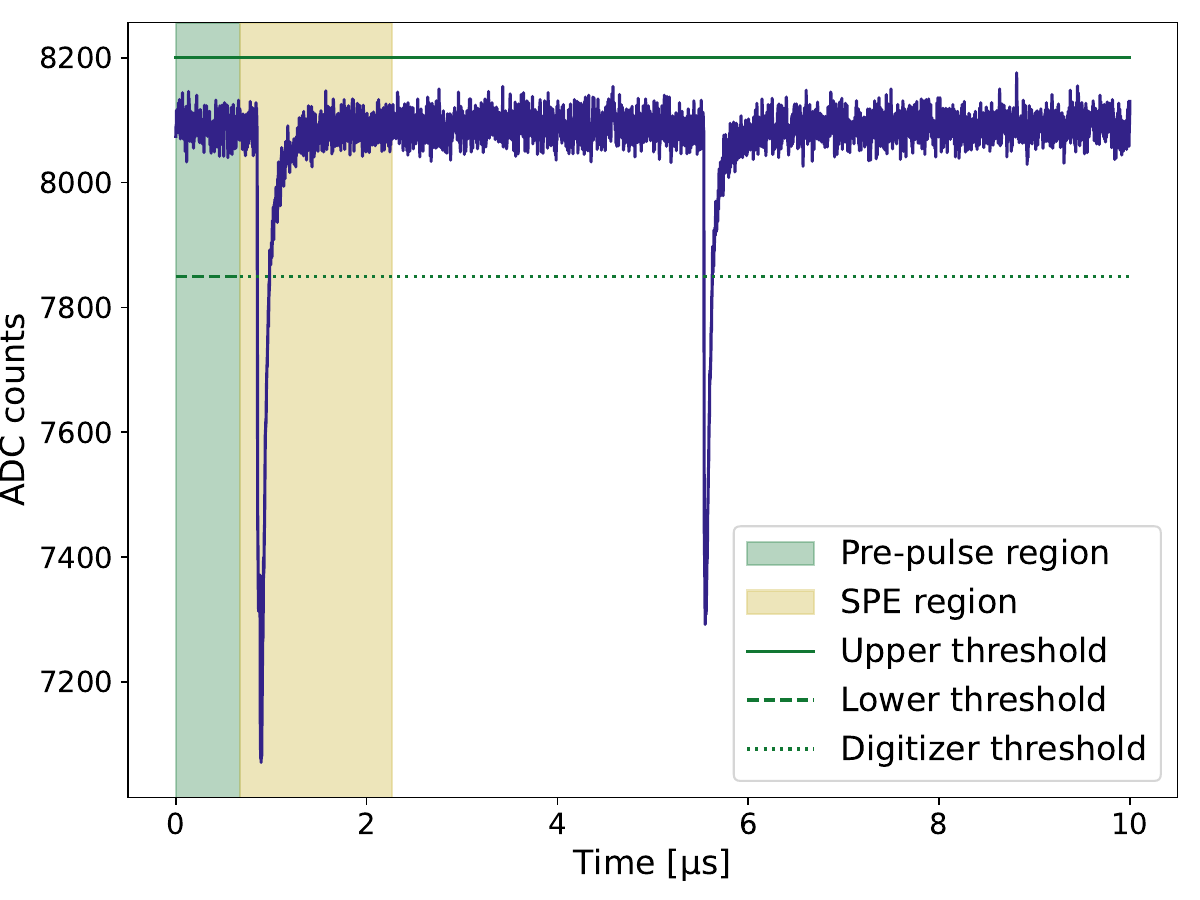}
    \caption{Waveform from the example SiPM at temperature 253.15~K and OV of 4~V. The pre-pulse region was used to calculate the baseline. The digitizer threshold is in units of ADC counts.}
    \label{fig:ana:example_waveform}
\end{figure}

The first step in the analysis was to clean up the waveforms to remove potentially ill-acquired signals. Preprocessing conditions required that the waveform must have no unusual features beyond the digitizer threshold in the pre-trigger region, the waveform must not have unusually high counts at any point, which for this setup could only be attributed to radio-frequency (RF) spurs, and finally, that the waveform is not empty (the total number of waveforms that pass these cuts has been defined as $N$). An example waveform with the regions described is shown in figure~\ref{fig:ana:example_waveform}.

After the waveforms have been pre-processed, the next step is to estimate   $g_{\text{SiPM}}$ which is then used to calculate $V_{\text{BD}}$. The estimation of $g_{\text{SiPM}}$ requires an algorithm that can efficiently identify SPE pulses from the dataset \cite{gallina2019}. When a set of SPE pulses is used, $g_{\text{SiPM}}$ is proportional to the average of the baseline-subtracted waveform areas. To increase the yield of SPE waveforms, the waveforms were restricted to a 1.6~$\upmu$s (or 800 samples) window after the trigger point as shown in the green region in figure~\ref{fig:ana:example_waveform}. This window length was selected to reduce summing error in the SPE calculation. Before calculating the sum, each individual waveform has its baseline removed, estimated as the mean of the pre-pulse region shown in the yellow area in figure~\ref{fig:ana:example_waveform}. 
% This analysis can work both for the ADC counts sum or the SiPM emitted charge. In this section, whenever the term ``charge" is used, it refers to the ADC counts sum. To avoid confusion in the notation, $A$ is used for the sum of the ADC counts and has arbitrary units, while $Q$ is exclusively reserved for the SiPM-generated charge in units of C and is calculated as $Q = \alpha A$. 

\begin{figure}[htbp]
    \centering
    \includegraphics[width=.55\textwidth]{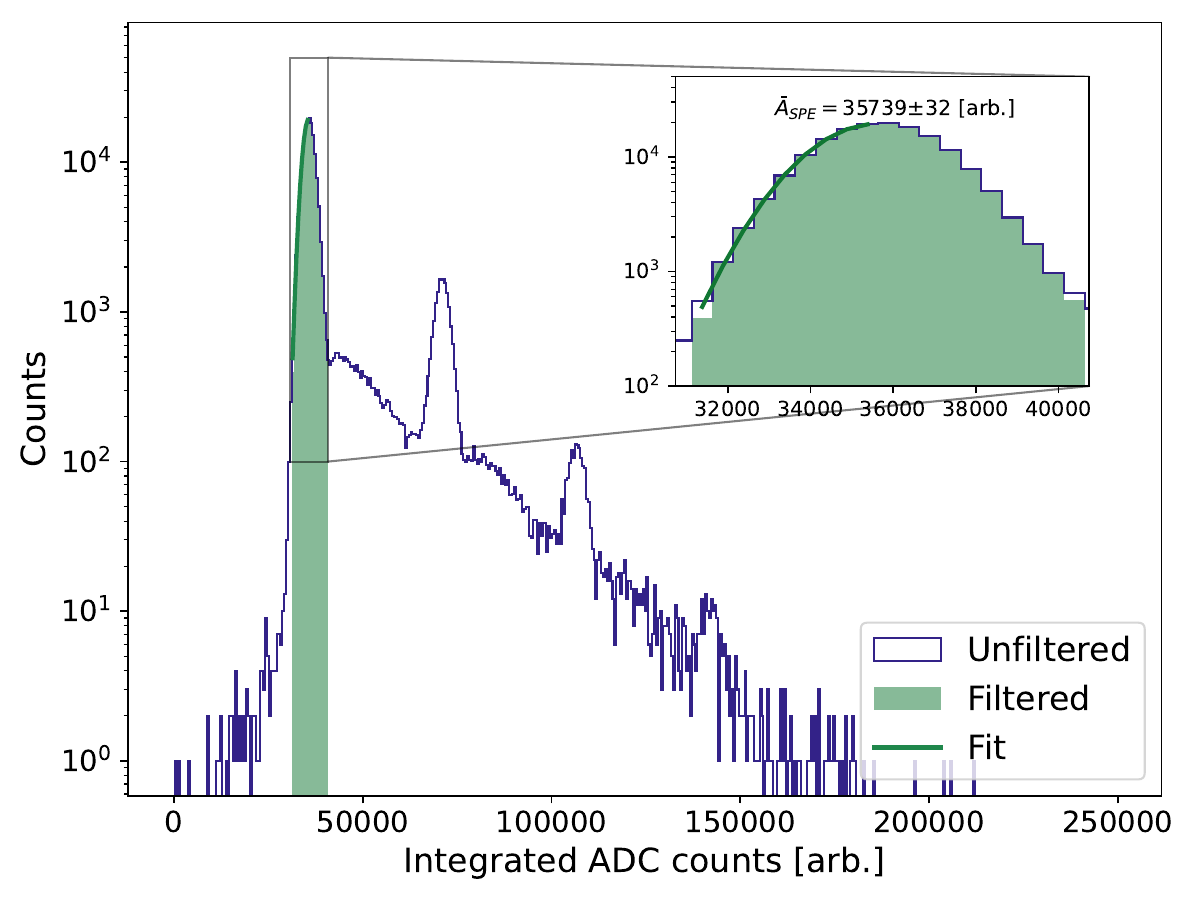}
    \caption{Finger plot of the example SiPM at temperature 253.15~K and OV of 4~V. The filtered region is the potential SPE cluster found from the sklearn DBSCAN algorithm \cite{pedregosa2011}. The SPE charge, $\bar{A}_{\text{SPE}}$, is estimated from a Gaussian fit to the left side of the potential SPE histogram.}
    \label{fig:ana:spe_estimation}
\end{figure}

Once the charge was calculated for all waveforms, the SPE was estimated from the most prominent peak in the charge histogram. The most reliable and efficient cluster-finding algorithm was found to be the DBSCAN algorithm from sklearn \cite{pedregosa2011}. DBSCAN finds the potential SPE cluster across all tested temperatures and voltages without any fine-tuning and with no observed failure rate. Once a DBSCAN calculation finishes, the estimated SPE charge, $\bar{A}_{\text{SPE}}$, is calculated from a Gaussian fit using the left side of the potential SPE charge histogram. This fit is a slightly better estimation of the SPE charge than the mean of the potential SPE charges. It avoids the portion of the potential SPE waveforms which might contain contributions from AP. Figure~\ref{fig:ana:spe_estimation} shows an example charge histogram or ``finger plot'', the potential SPE region returned from the DBSCAN calculation, and the fit. Alongside this step, 2,000 waveforms are randomly sampled from the potential SPE waveforms and used to calculate an average waveform which will be used during the ``wave-unfolding'' step. The SPE extraction step is repeated for all SiPMs at each voltage and temperature.

With $\bar{A}_{\text{SPE}}$ estimated for all the voltages and temperatures for a given cell, the average breakdown voltage $V_{\text{BD}}$ is estimated as a linear fit of the $\bar{A}_{\text{SPE}}$ vs.~voltage where the voltages are calculated from the averages of eq.~\ref{eq:sipm_voltage}.  Then over-voltage, OV, is calculated as OV$ = V - \hat{V}_{\text{BD}}$, and $m$ is calculated. $V_{\text{BD}}$ and $m$ are estimated for all SiPMs at both temperatures.

The timing parameter estimation requires an algorithm to extract the starting time and charge for each pulse found in a waveform. This type of algorithm can be called a ``charge fitting'' or ``wave-unfolding'' algorithm \cite{xu2022, peterson2021}. Currently, the time-dependent parameters commonly used in estimation use the time difference between the first independent pulse and the second pulse \cite{gallina2019, acerbi2019}. A more complex model of the time-dependent parameters is possible in which the first pulse is not assumed to be independent, and up to n-pulses can be used. However, correlations of CAs have to be accounted for in the final fit, which would lead to a complex model with many assumptions on the types of CAs, which depends on the internal physical mechanisms and geometry of the SiPM \cite{lgallego2013}. In this work, the wave-unfolding algorithm is a modification of \cite{peterson2021} that uses the average waveform calculated in the SPE estimation step instead of a pulse response function. The complete wave-unfolding algorithm is flexible, taking $\sim$10~ms for each waveform.

\begin{figure}[htbp]
    \centering
    \includegraphics[width=0.55\textwidth]{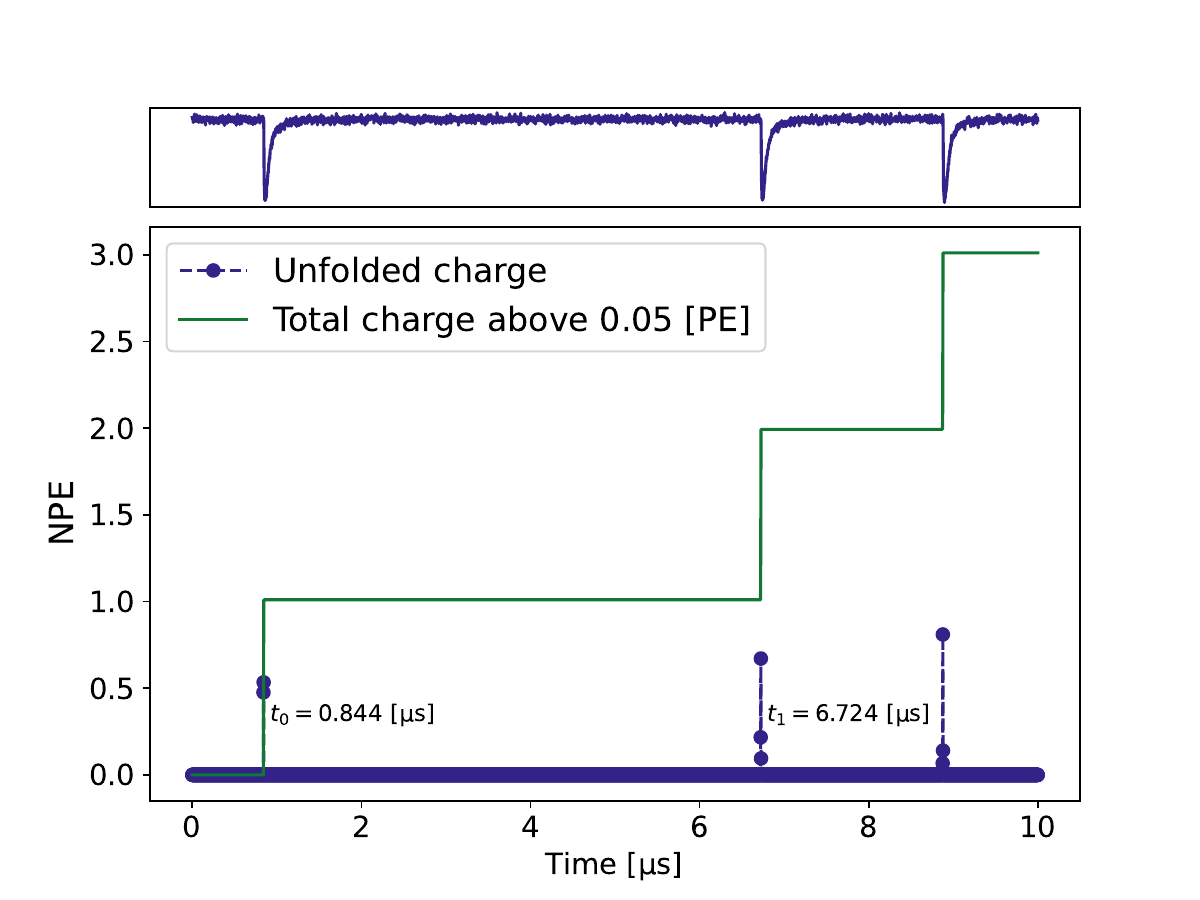}
    \caption{Output of the wave-unfolding algorithm developed for the extraction of the first and second pulse times and charges. The plot on the top is the raw waveform while the one below shows the unfolded charge and the total charge. The final unfolding window was limited to 4.5 $\upmu$s starting at the end of the pre-pulse region to reduce computation time.}
    \label{fig:ana:unfolding_example}
\end{figure}

To reduce computation time, the unfolding algorithm only extracted pulses in a 4.5 $\upmu$s window starting at the end of the pre-pulse region to reduce computation time while extracting pulses in the region of interest for SBC and for a DCR characterization. An example unfolded waveform with a 10 $\upmu$s acquisition window is shown in figure \ref{fig:ana:unfolding_example}.

\begin{figure}[htbp]
    \centering
    \includegraphics[width=0.55\textwidth]{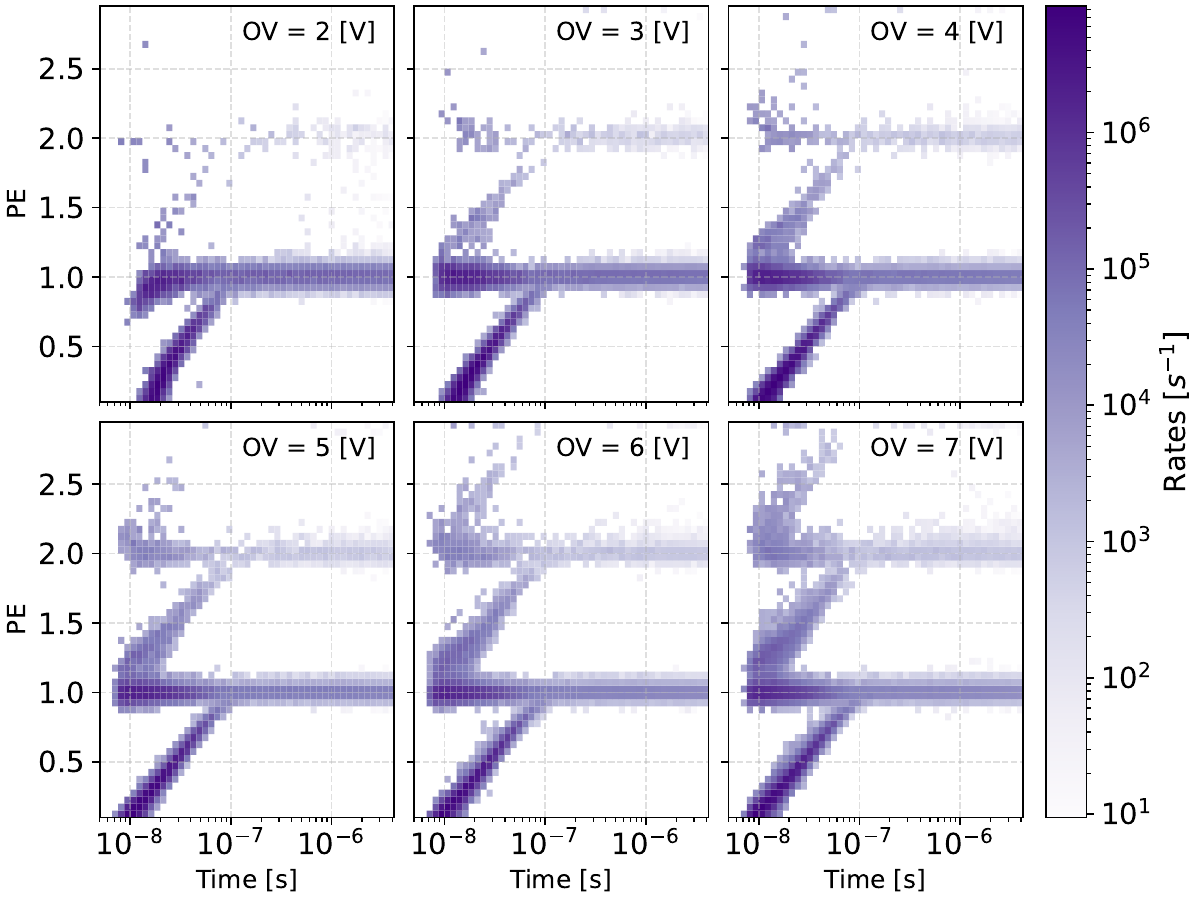}
    \caption{Timing and charge shadowed histograms for the SiPM with ID 200 and cell 1 at 253.15~K for 6 different OVs. The histogram for OV=8~V was left out for clarity. The time axis represents the time difference of the second pulse with respect to the first pulse.}
    \label{fig:ana:example_charge_time_hist}
\end{figure}

After the times and charges for the first and second pulses above 0.05~PE are extracted from the wave-unfolding algorithm, a 2D histogram is built from the time difference between the two pulses and the charge of the second pulse. Then, the 2-D histogram is collapsed in the time-axis, and the counts beyond the expected DCA window (150~ns) are used to estimate the SiPM DCR. The counts must be compensated by a bias introduced by the second pulse inhibiting any sequential pulse. For example, if two dark count pulses appear after the triggering pulse in the waveform, the first dark noise pulse will only be counted. This bias is called ``shadowing'', an effect unique to waveforms that might contain 3 or more pulses \cite{butcher2017}. The unshadowed counts, $k_i$, can be calculated using
\begin{equation}
    k_i = N \log{\left (\frac{N - n_i - \sum_{j=0}^i n_j}{N-\sum_{j=0}^i n_j}\right ) },
\end{equation}
where $n_i$ are the counts at bin $i$ with a bin width equal to $\Delta t_i$. Examples of the shadowed histograms for both charge and time with respect to the first pulse are shown in figure \ref{fig:ana:example_charge_time_hist}. The projected unshadowed time histogram compared to the projected histogram with no modifications can be seen in figure~\ref{fig:ana:example_charge_time_hist}. 
\begin{figure}[htbp]
    \centering
    \includegraphics[width=0.55\textwidth]{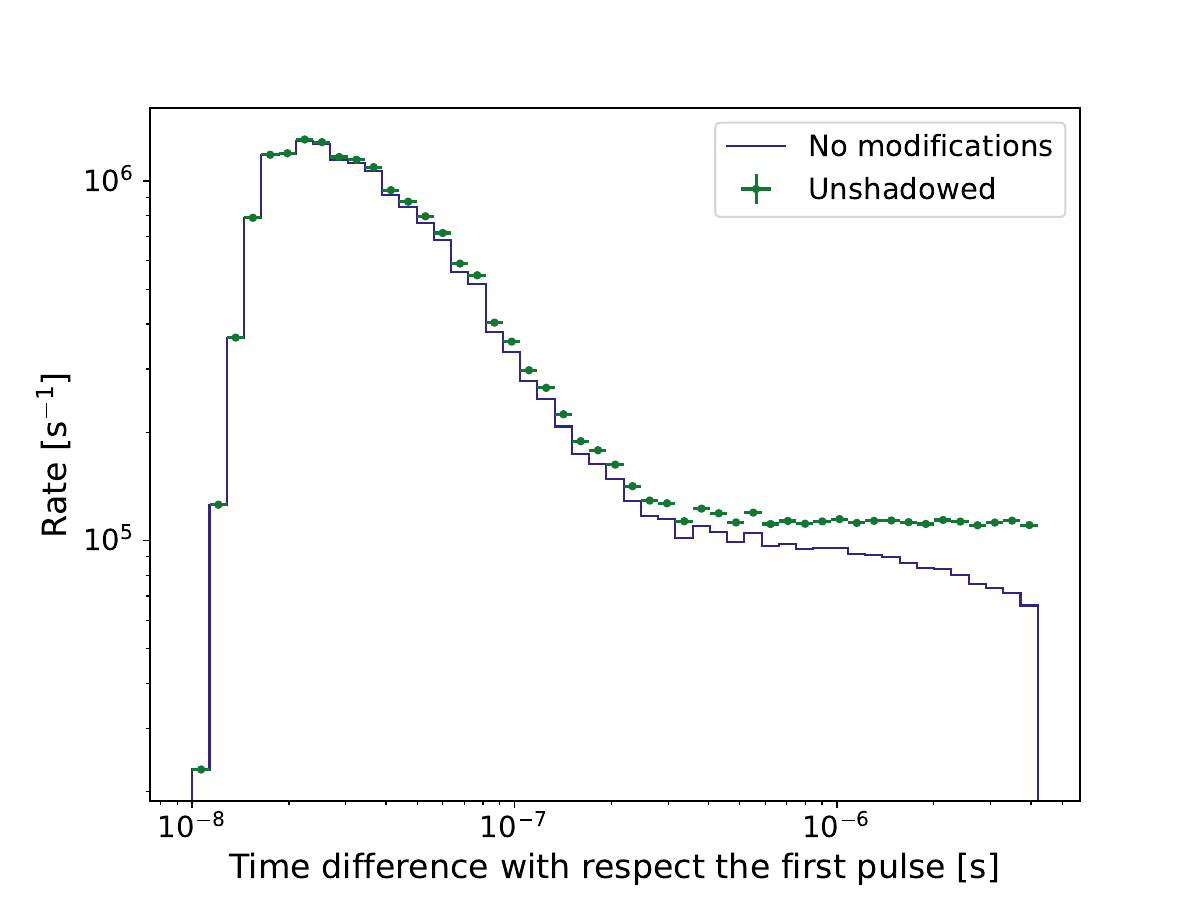}
    \caption{Time projected charge-time histogram for OV=4~V from Fig.~\ref{fig:ana:example_charge_time_hist}. Unshadowed values are shown with error bars. The flat portion on the right of the unshadowed values is equal to DCR.}
    \label{fig:ana:example_unshadowed}
\end{figure}

To simplify the estimations of DCR and P$_{\text{DCA}}$, the histogram is reduced to two-time bins, one bin ($\Delta t_0$) that contains all the DCA pulses plus a background portion of dark noise pulses, $n_0$, and the other bin ($\Delta t_1$) that contains only dark noise pulses, $n_1$. An estimator of DCR, $\widehat{\text{DCR}}$, can be calculated from
\begin{equation}
    \widehat{\text{DCR}} = \frac{k_1}{N\Delta t_1}.
\end{equation}

An estimator of P$_{\text{DCA}}$, $\hat{\text{P}}_{\text{DCA}}$, can be calculated as the ratio of the background-subtracted first unshadowed counts ($k_0$) with respect $N$
\begin{equation}
    \hat{\text{P}}_{\text{DCA}} = \frac{1}{N}\left (k_0 - \text{DCR} \frac{\Delta t_0}{\Delta t_1}\right ),
\end{equation}
and the errors for $\widehat{\text{DCR}}$ and $\hat{\text{P}}_{\text{DCA}}$ were estimated with standard error propagation. 

Using the set of independent first pulses, P$_{\text{PCA}}$ is estimated as the ratio of the number of independent first pulse charge counts $n'_i$ where the charge is higher than 1.5~PE and $N$. The 1.5~PE threshold is chosen to avoid SPE pulses with small AP. This can be written as
\begin{equation}
    \hat{\text{P}}_{\text{PCA}} = \frac{1}{N}\sum\limits_{Q > 1.5\text{PE}}n'_{j}.
\end{equation}
Then, $\hat{\text{P}}_{\text{CA}}$, under the assumption that $\text{{P}}_{\text{PCA}}$ and $\text{P}_{\text{DCA}}$ are independent which is true for low probabilities (see ref.~\cite{lgallego2013}), is approximately equal to
\begin{equation}
    \hat{\text{P}}_{\text{CA}} = \hat{\text{P}}_{\text{PCA}} + \hat{\text{P}}_{\text{DCA}}.
\end{equation}
Finally, $\bar{\tau}_{\text{DCA}}$ is the average of the DCA arrival times with respect to the first pulse. This can be calculated from a background-subtracted re-binned histogram of the time difference with respect to the first pulse with additional bins, $M$, for a better timing resolution. This can be expressed as 
\begin{equation}
    \bar{\tau}_{\text{DCA}} = \frac{\sum_{i}^M t_i ( k_i - N~\rm{DCR}~\Delta t_i)}{\sum_i^M k_i - N~\rm{DCR}~\Delta t_i}.
\end{equation}
All of the timing parameters are calculated for all SiPMs at each OV and both temperatures.

\section{Results}
\label{sec:results}
The analysis results for all 32 SBC VUV4 Hamamatsu Quads are presented in this section. The plots are labelled to distinguish between the two different Quad batches as discussed in section~\ref{sec:intro}, and to differentiate between the testing temperatures. The plots have the example SiPM discussed in section~\ref{sec:analysis} (Quad ID 200 and SiPM 1) visualized using semi-transparent grey lines whenever possible. Additionally, any single numbers extracted from the data, such as averages or temperature coefficients, will come in pairs. The first shows the SiPM-to-SiPM variations where the reported error is the standard deviation of the spread. The second value (labelled as $s_x$ where $x$ is the parameter of interest) is the average of the single SiPM errors which reflects the uncertainty of each single calculated value. The dispersion $V_{\text{BD}}$ and $m$ within the SiPMs in a Quad will also be discussed.

\subsection{SiPM-to-SiPM parameters spread}
\begin{figure}[htbp]
    \centering
    \includegraphics[width=0.55\textwidth]{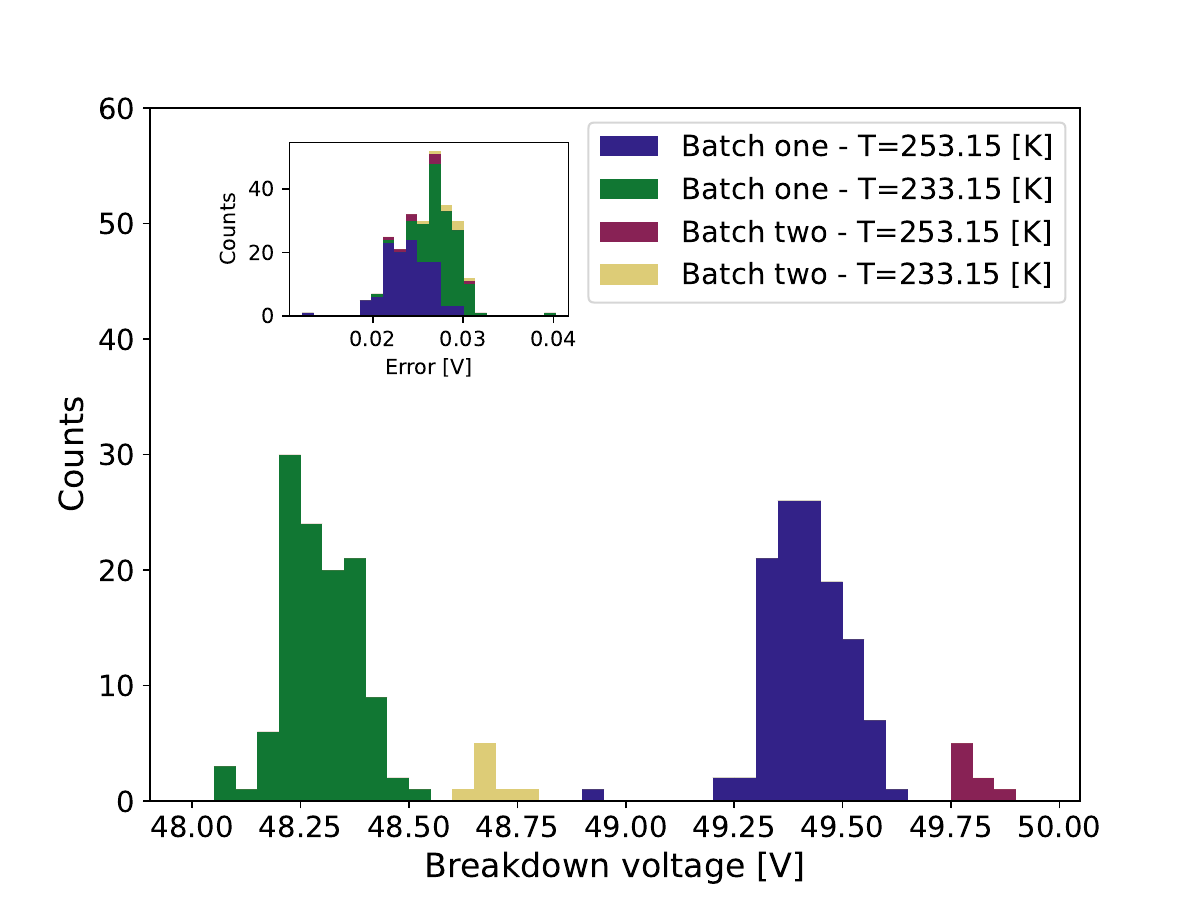}
    \caption{$V_{\text{BD}}$ was calculated from a linear fit of the charges vs.~voltage. The two batches show a different average value in the spread across both temperatures. The errors of the calculated $V_{\text{BD}}$ are shown in the inset histogram.}
    \label{fig:results:vbd}
\end{figure}

Shown in figure~\ref{fig:results:vbd}, the average value for $V_{\text{BD}}$ has been calculated for both batches, and they are detailed in table~\ref{table:vbds}. On average between both temperatures, $V_{\text{BD}}^{\text{Batch }1}$ is always smaller than $V_{\text{BD}}^{\text{Batch }2}$. This difference could be attributed to a slight difference in the manufacturing process, causing batch-to-batch variations only on long-term batch time differences. Another possible cause of the variations is storage damage. Despite the efforts to minimize humidity damage over the past several years since acquisition, humidity could have caused the $V_{\text{BD}}^{\text{Batch }1}$ to drift more in batch one than in batch two. The second option is less likely because $V_{\text{BD}}$ variations are from variations in the epitaxial layer resistivity that is dominated by manufacturing processes \cite{serra2011a}.
%For 253.15~K, the average $V_{\text{BD}}$ for batch one ($V_{\text{BD}}^{B1, T=253.15~\text{K}}$) is equal to $49.42\pm0.09$~V with a individual SiPM error, $s^{B1,T=253.15~\text{K}}$, equal to $0.024\pm0.003$~V. While for batch two $V_{\text{BD}}^{B2, T=253.15~\text{K}}=49.8\pm0.03$~V ($s^{B2, T=253.15~\text{K}}=0.028\pm0.002$~V). For 233.15~K, the average values are $V_{\text{BD}}^{B1,T=233.15~\text{K}}=48.29\pm0.08$~V ($s^{B1,T=233.15~\text{K}}=0.028\pm0.002$~V), and $V_{\text{BD}}^{B2,T=233.15~\text{K}}=48.68\pm0.03$~V ($s^{B2,T=233.15~\text{K}}=0.028\pm0.001$~V). 
\begin{table}[h!]
\centering
%\begin{tabular}{|l|l|l|}
%\hline
%& $V_{\text{BD}}$ {[}V{]} & $s_{V_{\text{BD}}}$ {[}V{]} \\ \hline
%\begin{tabular}[c]{@{}l@{}}Batch 1\\ T = 253.15 K\end{tabular} & $49.42\pm0.09$   & $0.024\pm0.003$     \\ \hline
%\begin{tabular}[c]{@{}l@{}}Batch 2\\ T = 253.15 K\end{tabular} & $49.80\pm0.03$   & $0.028\pm0.002$     \\ \hline
%\begin{tabular}[c]{@{}l@{}}Batch 1\\ T = 233.15 K\end{tabular} & $48.29\pm0.08$   & $0.028\pm0.002$     \\ \hline
%\begin{tabular}[c]{@{}l@{}}Batch 2\\ T = 233.15 K\end{tabular} & $48.68\pm0.03$   & $0.028\pm0.001$     \\ \hline
\begin{tabular}{|l|l|l|l|}
\hline
Temperature [K] & Batch  & $V_{\text{BD}}$ [V] & $s_{V_{\text{BD}}}$ [V] \\ \hline
\multirow{2}{*}{253.15} & one & $49.42\pm0.09$   & $0.024\pm0.003$     \\ 
                        & two & $49.80\pm0.03$   & $0.028\pm0.002$     \\ \hline
\multirow{2}{*}{233.15} & one & $48.29\pm0.08$   & $0.028\pm0.002$     \\ 
                        & two & $48.68\pm0.03$   & $0.028\pm0.001$     \\ \hline
\end{tabular}
\caption{Average $V_{\text{BD}}$ across batches and temperatures. $s_{V_{\text{BD}}}$ is the average error of each individual SiPM $V_{\text{BD}}$.}
\label{table:vbds}
\end{table}

The rate of change of the $V_{\text{BD}}$ with respect to temperature, $\Delta V_{\text{BD}}$, is $56 \pm 2$~mV~K$^{-1}$. The single SiPM error average of $\Delta V_{\text{BD}}$ is $s_{\Delta V_{\text{BD}}}=1.8\pm0.1$~mV~K$^{-1}$. The error of $\Delta V_{\text{BD}}$ is dominated by the SiPM-to-SiPM variations while $s_{\Delta V_{\text{BD}}}$ is on average 30\% smaller than the error of $V_{\text{BD}}$. 
%On average, it is more accurate to predict a single SiPM $V_{\text{BD}}$ across temperature once the $V_{\text{BD}}$ has been estimated for at least two temperatures.

\begin{figure}[htbp]
    \centering
    \includegraphics[width=0.55\textwidth]{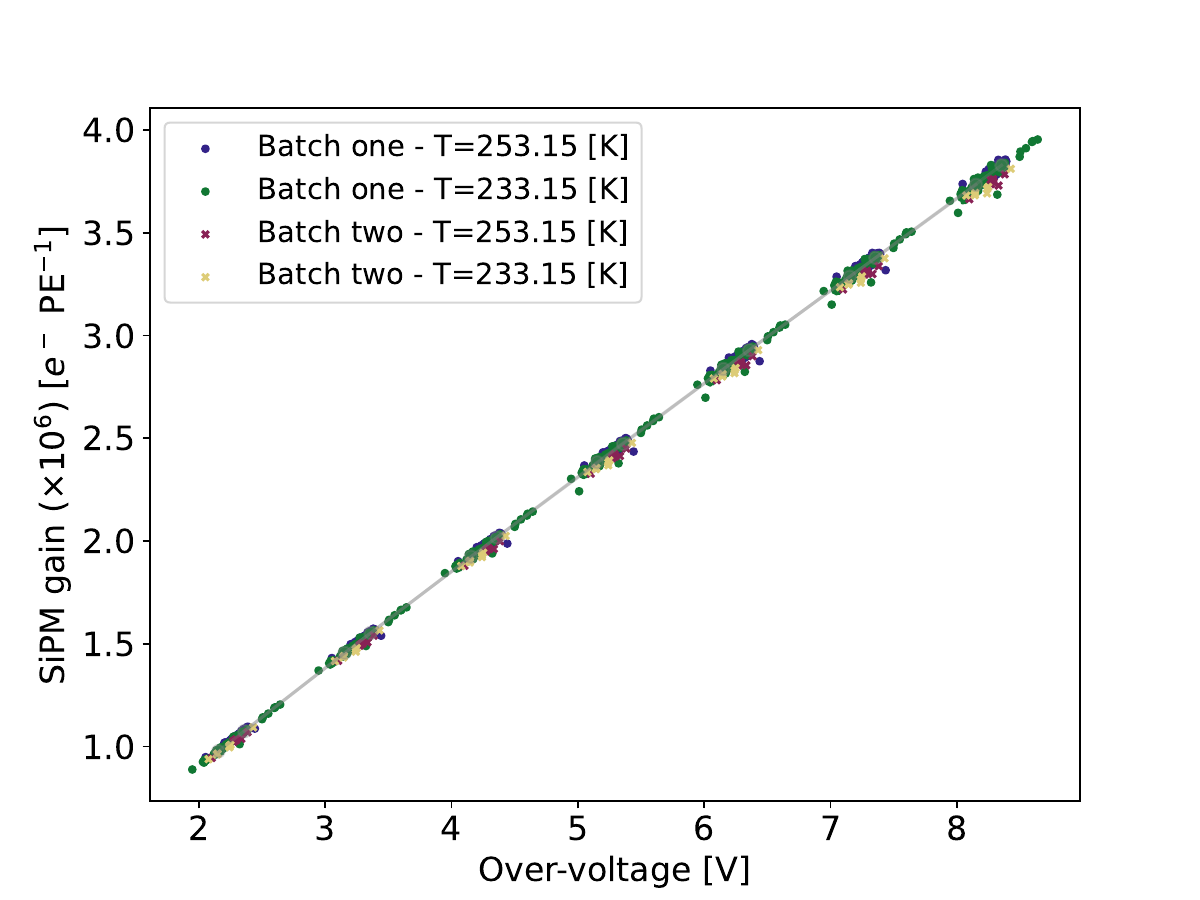}
    \caption{The SiPM gain, $g_{\text{SiPM}}$, as a function of OV. The gains were calculated with eq.~\ref{eq:sipm_gain} from the estimated $\bar{A}_{\text{SPE}}$. Note that Quads 218 and 220 had the gain correction applied for data taken at 233.15~K, and batch two had the correction applied for all temperatures.}
    \label{fig:results:gain}
\end{figure}

Once $V_{\text{BD}}$ was estimated for each SiPM, OV could be calculated for each data point. The $g_{\text{SiPM}}$ calculated from eq.~\ref{eq:sipm_gain} was plotted against the calculated OV and is shown in figure~\ref{fig:results:gain}. The $g_{\text{SiPM}}$ vs.~OV plot shows that the spread of the $g_{\text{SiPM}}$ with respect to OV is relatively minor and has no dependence on temperature. An explanation of the higher $V_{\text{BD}}$ in batch two can be found in this figure as, in general, batch two has lower gains, especially observable at higher OVs. 

A feature left out of figure~\ref{fig:results:gain} is that a single SiPM was found to have unusually high OVs (up to 11~V). When cross-checking the same SiPM in figure~\ref{fig:results:vbd}, it was observed that the single count from batch one at $T=252.15$~K near the 49~V bin belongs to the same SiPM. After analyzing the shape of the unusual SiPM finger plot, this SiPM had a rare type of pulse that was half the height of a normal SPE pulse with different rise and fall times. These unusual pulses skewed the analysis to lower gains which would be reflected in a lower $V_{\text{BD}}$.

\begin{figure}[htbp]
    \centering
    \includegraphics[width=0.55\textwidth]{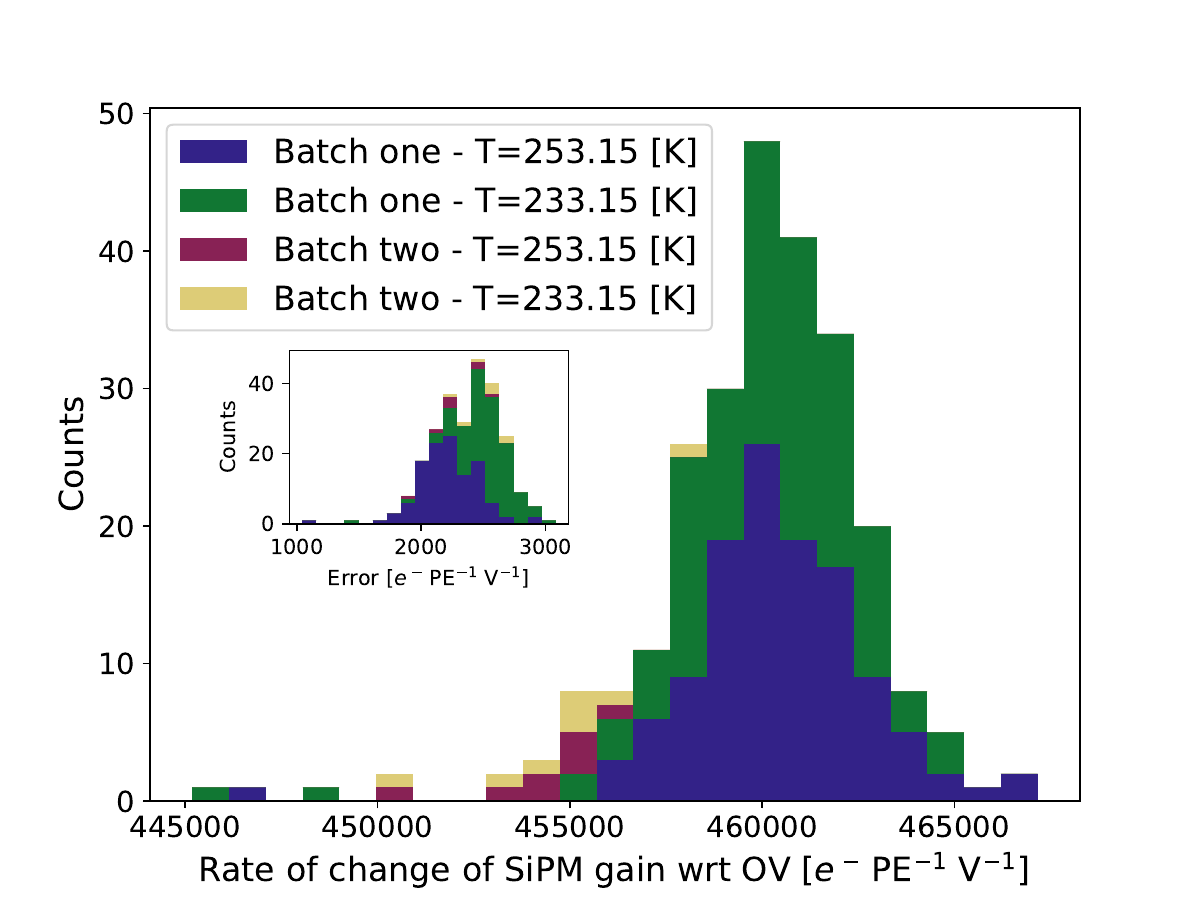}
    \caption{Rate of change of the SiPM gain with respect to OV, $m$, histogram. The inset error histogram shows only statistical error. Note that SiPMs 218 and 220 had the correction applied for data taken at 233.15~K, and batch two had the correction applied for all temperatures.}
    \label{fig:results:slope}
\end{figure}

The slope, $m$, was calculated alongside $V_{\text{BD}}$ using the estimated $\bar{A}_{\text{SPE}}$, and eq.~\ref{eq:sipm_gain} was used to get the slope in units of SiPM gain with the applicable corrections. The histogram of $m$ is shown in figure~\ref{fig:results:slope}. Similar to $V_{\text{BD}}$, there is a difference in the mean value between the two batches, however, due to a larger spread on the distribution of the slope, they overlap. There is no apparent temperature dependence of $m$ across batches. The average $m$ across all SiPMs is equal to $(459\pm3(\rm{stat.})\pm23(\rm{sys.}))\times 10^{3}~e^-$~PE$^{-1}$~V$^{-1}$ ($s_m = 2400 \pm 300$~$e^-$~PE$^{-1}$~V$^{-1}$). The systematic uncertainty comes from the artificial 5\% included as the error of $\alpha$ in eq.~\ref{eq:sipm_gain} discussed in section~\ref{sec:exp_setup}.

\begin{figure}[htbp]
    \centering
    \includegraphics[width=\textwidth]{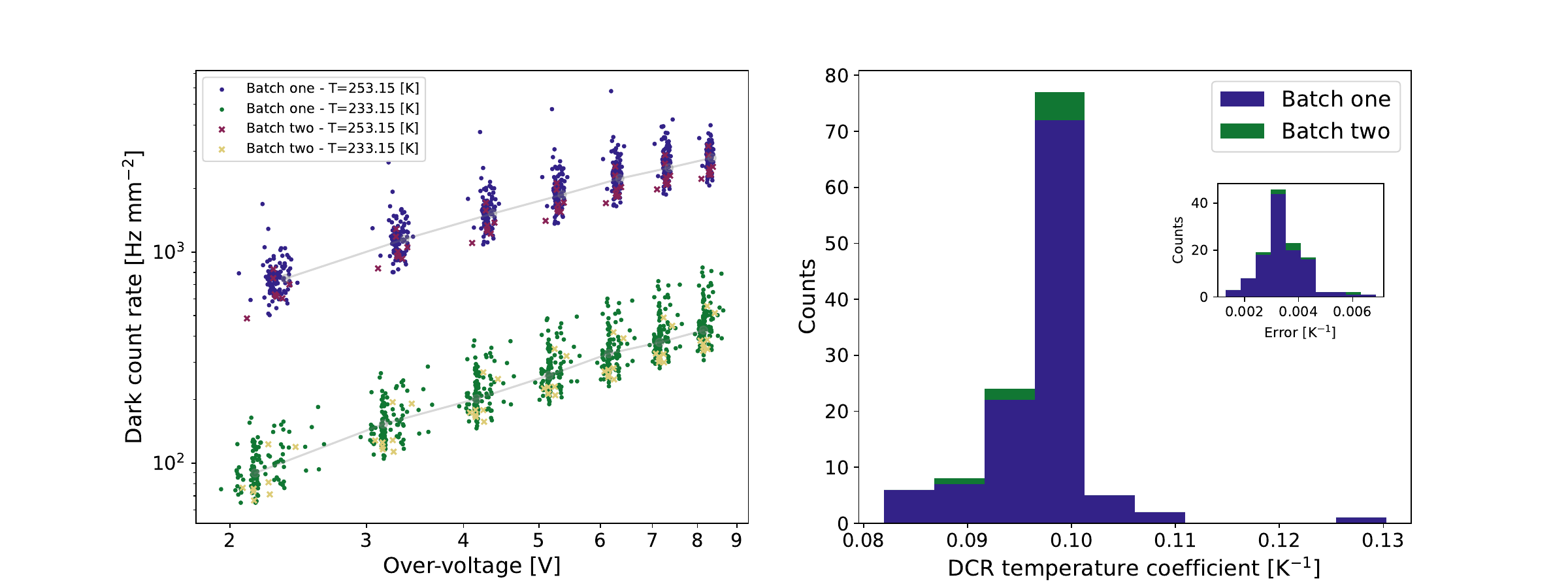}
    \caption{Left: DCR spread as a function of temperature for all SiPMs. This plot shows the temperature dependence of DCR decreases by a factor of 7 per 20~K across all OVs. Right: histogram of the calculated DCR temperature coefficients.}
    \label{fig:results:DCR}
\end{figure}

The first timing parameter plot is the DCR as a function of OV shown on the left side of figure~\ref{fig:results:DCR}. In contrast to the previous plots, the DCR shows a very strong temperature dependence. To extract the temperature coefficient of DCR, for each SiPM, a power law was fit to its $T=233.15$~K data. After fixing the exponent, the fit was repeated for the $T=253.15$~K data. Then, the log ratio of the coefficients was calculated and then divided by the temperature difference. The average value of the DCR temperature coefficient across all SiPMs was calculated to be $0.099\pm0.008$~K$^{-1}$. The average error for each SiPM was calculated to be $0.0034\pm0.0008$~K$^{-1}$ which includes a systematic of the power law failing at the $T=253.15$~K data for high OVs as explained in appendix~\ref{app:mc_sipm}. The differences between the standard deviation and individual SiPM error in the DCR temperature coefficient are attributed to the SiPM-to-SiPM variations, not an unaccounted systematic.

\begin{figure}[htbp]
    \includegraphics[width=\textwidth]{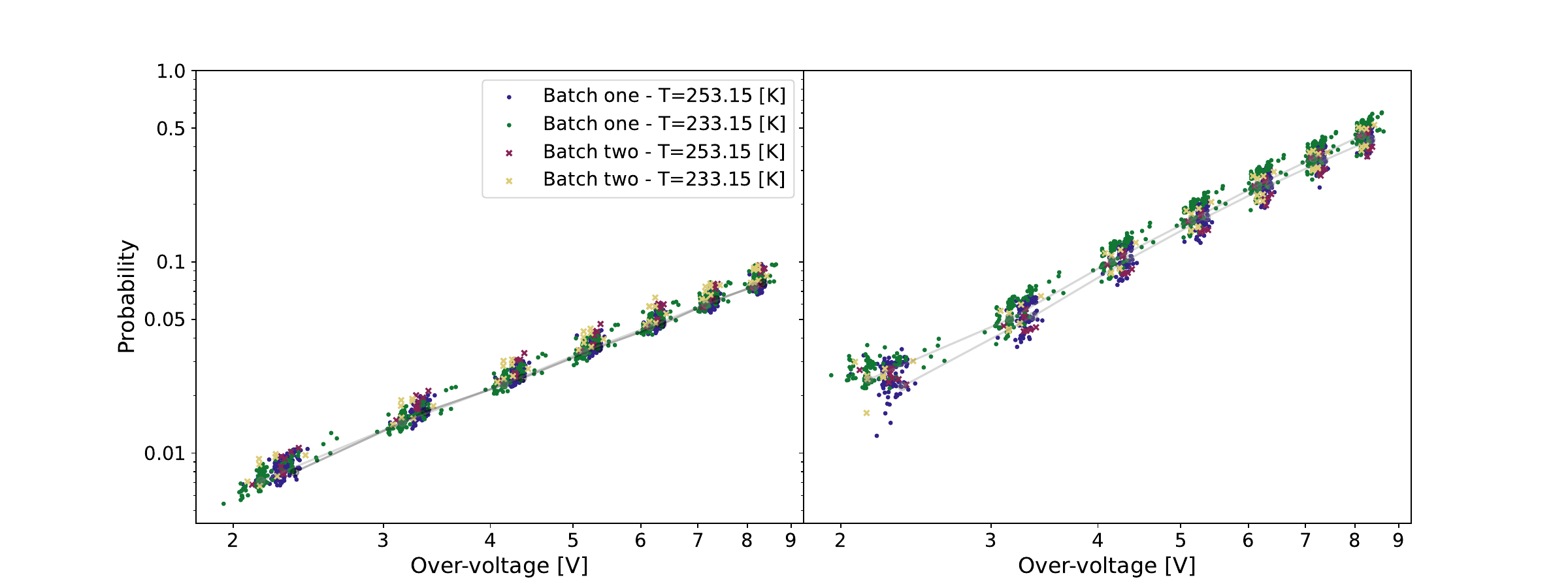}
    \caption{The spread in probability of a prompt correlated avalanche, P$_{\text{PCA}}$, (left) and of delayed correlated avalanches, P$_{\text{DCA}}$, (right) as a function of OV. Note the x-axis is in log scale.}
    \label{fig:results:P}
\end{figure}

\begin{figure}[htbp]
    \includegraphics[width=\textwidth]{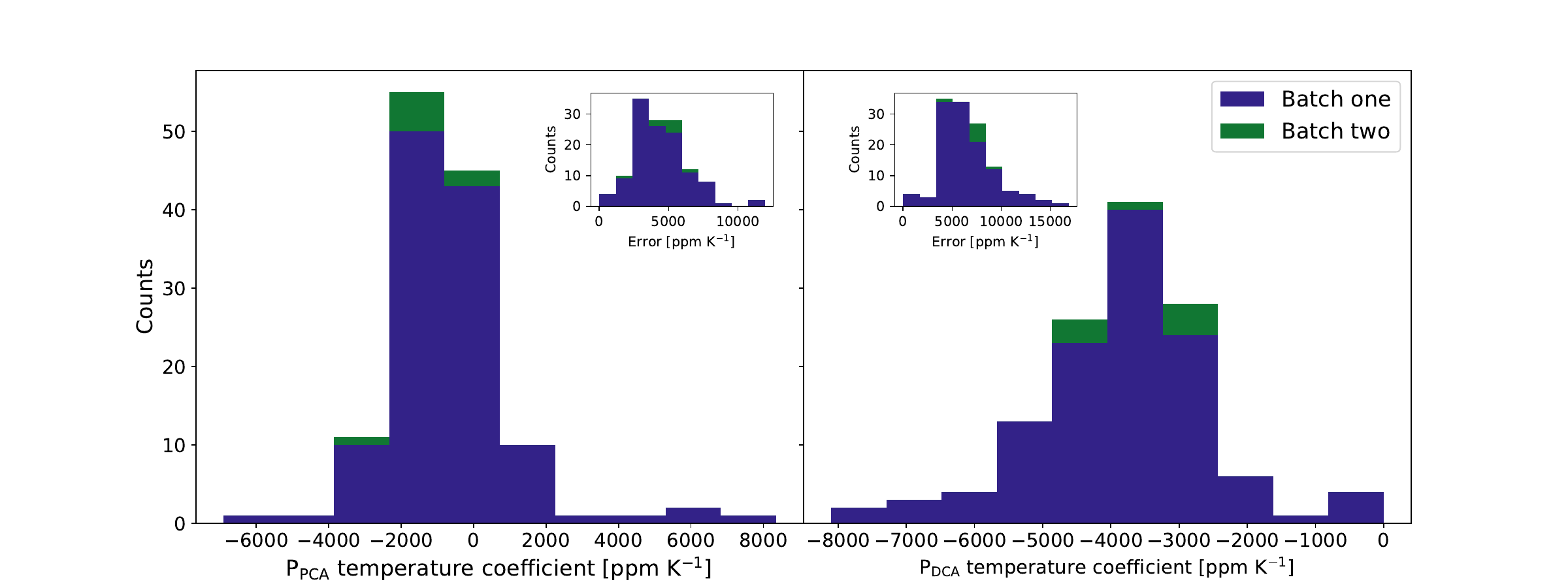}
    \caption{Temperature coefficients of P$_{\text{PCA}}$ (left) and P$_{\text{DCA}}$ (right).}
    \label{fig:results:P_tempco}
\end{figure}

The probabilities of the different types of CAs are shown in figure~\ref{fig:results:P}. The temperature coefficients for PCA and DCA were also calculated and their distributions are shown in figure~\ref{fig:results:P_tempco}. The P$_{\text{PCA}}$ was found to have an insignificant dependence on temperature calculated to be $-300\pm1400$~ppm~K$^{-1}$ ($s_{\text{P}_\text{PCA}}=3000\pm2000$~ppm~K$^{-1}$). While P$_{\text{DCA}}$ has a weak temperature dependence of $4000\pm1000$ ppm~K$^{-1}$ ($s_{\text{P}_\text{DCA}}=5000\pm2000$~ppm~K$^{-1}$). In contrast with the DCR temperature coefficient or the spread of $V_{\text{BD}}$, the single SiPM temperature coefficient errors are larger than the standard deviation. The consequence is that predicting the probabilities of a CA as a function of temperature for a single SiPM is not reliable.

\begin{figure}[htbp]
    \centering
    \includegraphics[width=0.55\textwidth]{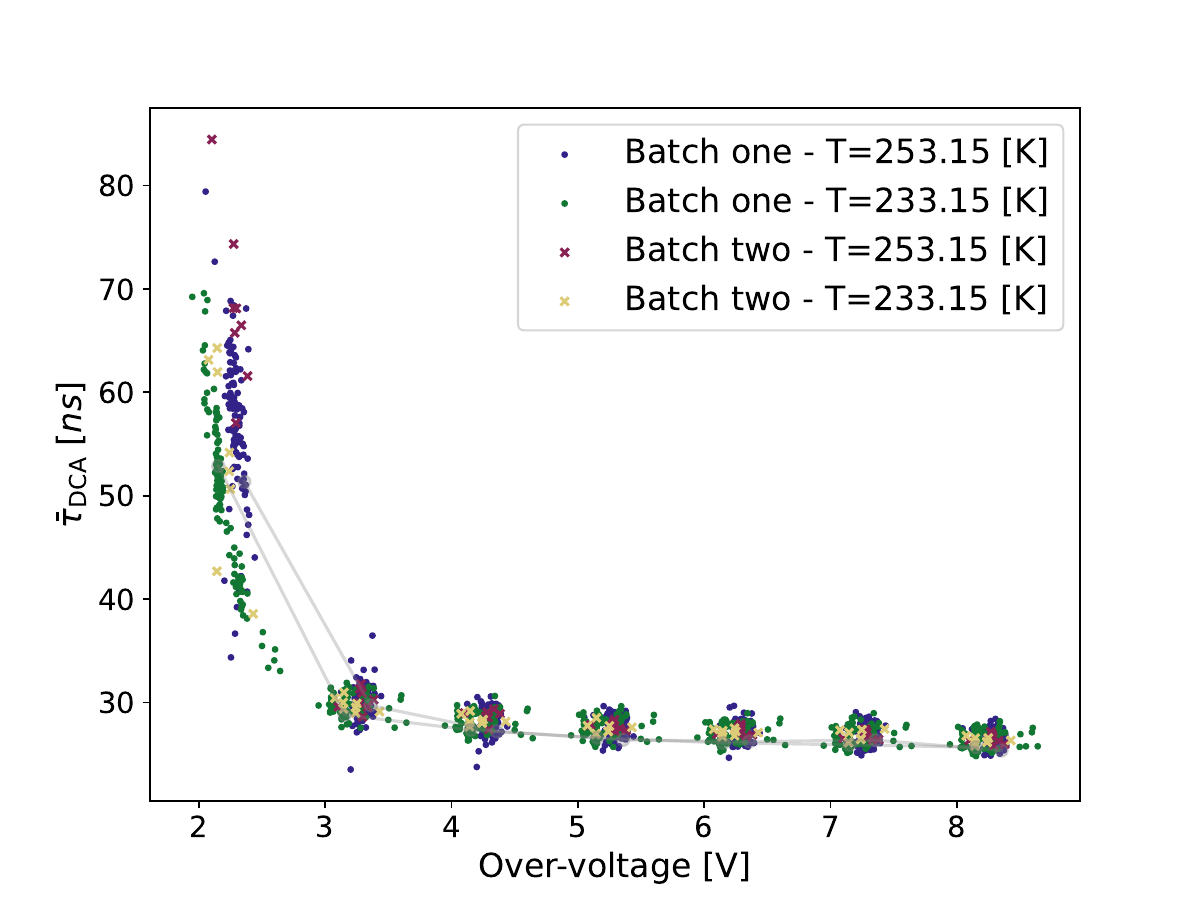}
    \caption{The average arrival time of delayed correlated avalanches, $\bar{\tau}_{\text{DCA}}$, as a function of OV.}
    \label{fig:results:tau}
\end{figure}

The final timing plot is for $\bar{\tau}_{\text{DCA}}$ shown in figure~\ref{fig:results:tau}. At an OV of 2~V, the data are highly correlated with OV and are higher compared to the rest of the data. The higher than usual $\bar{\tau}_{\text{DCA}}$ can be explained from the figure~\ref{fig:ana:example_charge_time_hist} OV = 2~V time-charge histogram. Low counts can be seen at low times on the left side of the AP tail. No physical mechanism is known, but the lack of this feature in the MC SiPM excludes the origin at the analysis level (see appendix~\ref{app:mc_sipm}). No temperature coefficient is calculated for this parameter. 

% The calculated $\bar{\tau}_{\text{DCA}}$ is the average of several DCA mechanisms with different time scales\cite{}. Additionally, this methodology assumes there is no DCA mechanism longer than the millisecond range.

\subsection{Quad-to-Quad $V_{\text{BD}}$ and $m$ dispersion}
\begin{figure}[htbp]
    \centering
    \includegraphics[width=\textwidth]{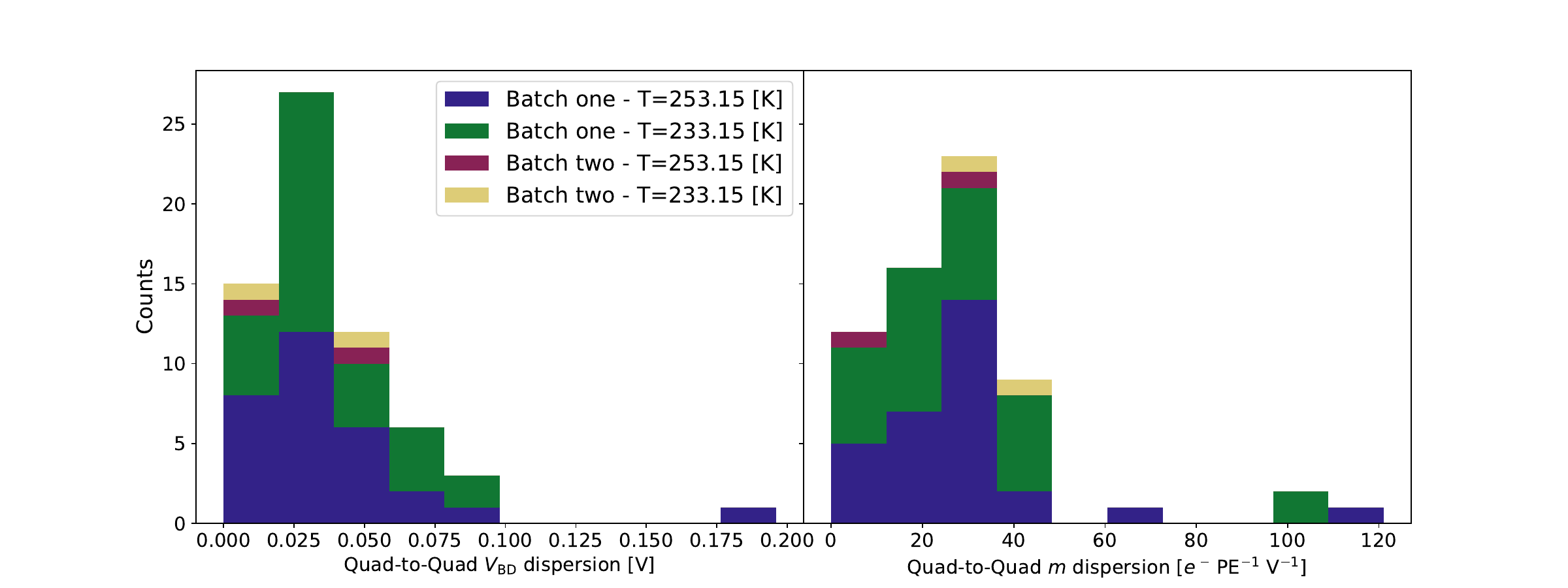}
    \caption{Left: $V_{\text{BD}}$ Quad-to-Quad dispersion. The count near 0.2~V is the Quad with the SiPM with an unusual pulse response. Right: $m$ Quad-to-Quad dispersion.}
    \label{fig:results:quadtoquad}
\end{figure}

The spread of $V_{\text{BD}}$ and $m$ within each Quad is relevant to SBC as the SiPMs will be ganged together to save on electronics channels. Connecting the Quads in parallel, however, might introduce uncertainty in $g_{\text{SiPM}}$ as the four SiPMs $V_{\text{BD}}$ and $m$ might not be paired. The dispersion of $V_{\text{BD}}$ and $m$ in each Quad was calculated as the standard deviation of each parameter within a Quad and are shown in figure~\ref{fig:results:quadtoquad}. 

The most notable observation from the left side of figure~\ref{fig:results:quadtoquad} is that the $V_{\text{BD}}$ dispersion is within the margin of error for the majority of cases. There is a very small Quad-to-Quad dispersion (less than 100~$e^-$~PE$^{-1}$~$V^{-1}$) in $m$ compared to the total dispersion (3000~$e^-$~PE$^{-1}$~$V^{-1}$). Both of these parameters are indicators that Hamamatsu pairs each SiPM in a Quad based on $V_{\text{BD}}$ and $m$. However, in rare cases, the dispersion is significant, and the consequence is that the Quad gain will have an additional source of error from different gains for a given voltage for each SiPM. For example, a $V_{\text{BD}}$ dispersion of 75~mV multiplied by the average $m$ across all SiPMs leads to a gain (and therefore pulse-height) dispersion of approximately 2\% at an OV of 4~V.

\section{Conclusions}
\label{sec:conclusions}
The setup and measurement procedure presented in this paper met the required conditions for the characterization with low-temperature variations and no significant errors using dark count pulses. The analysis was a reliable estimator of all parameters within the manufacturer distribution. In future characterization setups, the duration of data acquisition and analysis could be a point for improvement. In this setup, two Quads took eight hours for a full data acquisition routine, and each SiPM took 2 days for the analysis to complete. An improvement would be to scale up the SiPM box to hold more than two Quads at a time. The longest time bottleneck during the data acquisition was found to be waveform collection which was limited by the single-channel amplifier. More amplifiers would have reduced the waveform acquisition time to approximately 1~hr in length, reducing the total acquisition time to 4~hrs. During the analysis, most of the time was taken by the wave-unfolding algorithm. Future iterations could be improved by replacing it with a faster algorithm or using multi-processing routines such as GPU utilization. If the time consumed during the wave-unfolding is reduced to a few minutes, the analysis will take less than an hour per SiPM. 

The SBC SiPM implementation will use the estimated individual $V_{\text{BD}}$, temperature coefficient of $V_{\text{BD}}$, and $m$. The goal will be to set the SiPMs parameters for an optimal signal-to-noise ratio while keeping P$_{\text{CA}}$ insignificant for the dark matter search. A possible alternative scenario would be to use the SiPMs in a high OV to increase the signal-to-noise ratio without saturation as described in appendix~\ref{app:mc_sipm}. This will allow SBC to do energy reconstruction using the calculated P$_{\text{CA}}$ while increasing the DCR and the error in the energy estimator \cite{gallina2019}.

It was found that the $\bar{\tau}_{\text{DCA}}$ contribution to the pre-trigger length was insignificant at any OV as it is smaller than the pulse falling tails (in this setup is approximately equal to 1~$\upmu$s). Technologies are being developed to actively remove the falling tail of a pulse or actively remove the analog component of the signal at the SPAD level \cite{liu2022}. Therefore, future applications could be limited by CA which can be characterized by the methodology described in this paper.

The SiPM-to-SiPM $V_{\text{BD}}$ spread is roughly by a factor of four smaller (using full width at half maximum or FWHM as the estimator) than those found by MEG II at room temperature \cite{ieki2019}. MEG II had a bigger sample size (16720 sensors compared to 128 in this paper), and the spread includes batch-to-batch variations. If we compared the range of the spread of this paper at one temperature across both batches ($\sim0.75$~V) to the FWHM of MEG II ($\sim0.75$~V), they are similar. The Quad-to-Quad spread is also four times smaller than those of MEG II ($\sim0.1$~V compared to $\sim0.025$~V as estimated using the FWHM). 

This bulk characterization also shows that an individual SiPM characterization is a powerful tool for the estimation and prediction of the $V_{\text{BD}}$, $g_{\text{SPE}}$, and DCR at any temperature and OV. However, the average SiPM-to-SiPM P$_{\text{CA}}$ and P$_{\text{CA}}$ temperature coefficients are a more reliable estimator for any SiPM as the average of the errors is bigger than the error of the mean as shown in figure~\ref{fig:results:P_tempco}. For any other parameter, the error of the mean is bigger than the individual SiPM parameter error due to the significant contributions of the manufacturer spread to the average error. The DCR and DCR temperature coefficients can also be used to estimate the required temperatures for the desired application. The design of the SiPM implementation has to account for the significant drop of temperature dependence which for Hamamatsu VUV4 happens roughly around 150~K \cite{gallina2019}. Additionally, in liquid noble element scintillation experiments (90~K for LAr and 165~K for liquid xenon), the scintillation background and optical cross-talk will likely dominate over DCR.

\acknowledgments
% This work was supported by the Arthur B. McDonald Canadian Astroparticle Physics Research Institute and the Canada Foundation for Innovation (CFI), with additional support from the Natural Sciences and Engineering Research Council of Canada (NSERC). The Hamamatsu VUV4 SiPMs were purchased using resources provided by the Fermi National Accelerator Laboratory (Fermilab), a U.S. Department of Energy, Office of Science, HEP User Facility.
This document was prepared by the Scintillating Bubble Chamber (SBC) collaboration using the resources of the Fermi National Accelerator Laboratory (Fermilab), a U.S. Department of Energy, Office of Science, HEP User Facility. Fermilab is managed by Fermi Research Alliance, LLC (FRA), acting under Contract No. DE-AC02-07CH11359. The SBC collaboration also wishes to thank SNOLAB and its staff for support through underground space, logistical and technical services. SNOLAB operations are supported by the Canada Foundation for Innovation and the Province of Ontario Ministry of Research and Innovation, with underground access provided by Vale at the Creighton mine site. The SBC-LAr10 chamber at Fermilab and its twin at SNOLAB have been supported primarily by the Fermilab Lab-Directed R\&D program and the Canada Foundation for Innovation (CFI), respectively, with additional support from the Natural Sciences and Engineering Research Council of Canada (NSERC), the DOE Office of Science Graduate Instrumentation Research Award fellowship, the National Science Foundation under Grant No. PHY-2310112, the Arthur B. McDonald Canadian Astroparticle Physics Research Institute, the projects CONACyT CB2017-2018\/A1-S-8960, DGAPA UNAM grant PAPIIT IN105923, Fundaci\'on Marcos Moshinsky, the Indiana University South Bend Office of Research, the URA Visiting Scholar program, the Fermilab Cosmic Physics Center, and DOE Office of Science grants DE-SC0015910, DE-SC0017815, DE-SC0024254, and DE-SC0011702.

\appendix

\section{MC SiPM results}
\label{app:mc_sipm}
Testing the bias introduced by the analysis required a Monte-Carlo (MC) SiPM data set with known parameters. The MC data set was generated using \cite{garutti2021} as a framework. The parameters tested were $V_{\text{BD}}$, $g_{\text{SPE}}$, DCR, and P$_{\text{CA}}$. Each file in the MC data set was created to mimic the real data whenever possible (for example, the number of waveforms and the time length of each waveform). The most complex aspects of the waveforms were approximated. For example, the noise figures of each waveform were sampled from a Gaussian with a mean of 0 and a standard deviation approximately equal to the one found in the real setup. The baseline of each waveform is constant and at approximately the same level as the real data. The expected values for all the parameters in the MC data were extracted from the test SiPMs and ref.~\cite{gallina2019} modified to fit the MC model. No modifications were made to the analysis described in section~\ref{sec:analysis} to test the MC data.

\begin{figure}
    \centering
    \includegraphics[width=0.55\textwidth]{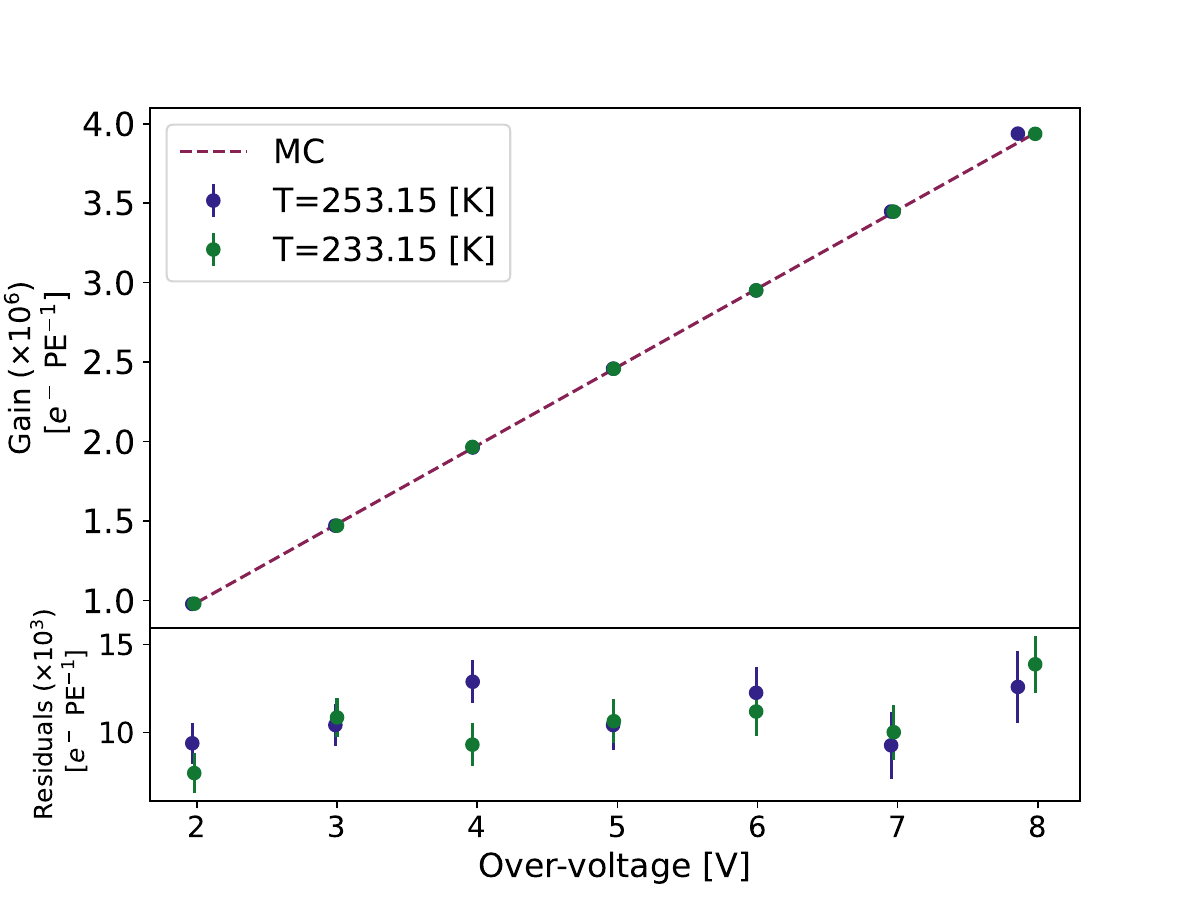}
    \caption{The gains of the MC SiPM as calculated from the analysis as a function of OV. The bias errors of $V_{\text{BD}}$ with respect to MC $V_{\text{BD}}$ at 253.15~K and 233.15~K are $0.04\pm0.02$~V and $0.01\pm0.01$~V. The $m$ bias error at both temperatures are $-750\pm2200~e^-$~PE$^{-1}$ and $-470\pm1118~e^-$~PE$^{-1}$. The data point for 253.15~K at 8~V has its voltage incorrectly estimated because of attributed to a miscategorization of a voltage point. Errors are statistical only.}
    \label{fig:mc:gain}
\end{figure}

Figure~\ref{fig:mc:gain} shows $g_{\text{SPE}}$. The gain was dependent only on OV, which had the same $m$ between both temperatures. The voltage and currents are drawn from a Gaussian with a mean equal to the expected OV, and their dispersion has been made ten times larger compared to the real data to study possible impacts of voltage fluctuations. Still, the impacts remained insignificant as long as they remained below the accuracy of the multimeter. Under these simplifications, any differences between both temperatures could only be attributed to DCR and the MC voltage data. As shown at the bottom of figure~\ref{fig:mc:gain}, the residuals of the gains show a bias for slightly overestimating the SPE gain. No definite cause was found for the over-estimation but the leading hypothesis is error attributed to the SPE window and baseline subtraction. The breakdown voltages and rates were calculated and the $V_{\text{BD}}$ bias errors with respect to MC $V_{\text{BD}}$ at 253.15~K and 233.15~K are $0.04\pm0.02$~V and $0.01\pm0.01$~V. The first one is not consistent with zero, which is attributed to a miscategorization of a voltage point in the voltage data at 8~OV while the other one is consistent. The $m$ bias error for both temperatures are $-750\pm2200~e^-$~PE$^{-1}~V^{-1}$ and $-500\pm1100~e^-$~PE$^{-1}~V^{-1}$. They are both consistent with 0 which is expected.

\begin{figure}
    \centering
    \begin{subfigure}{0.45\textwidth}
        \includegraphics[width=\textwidth]{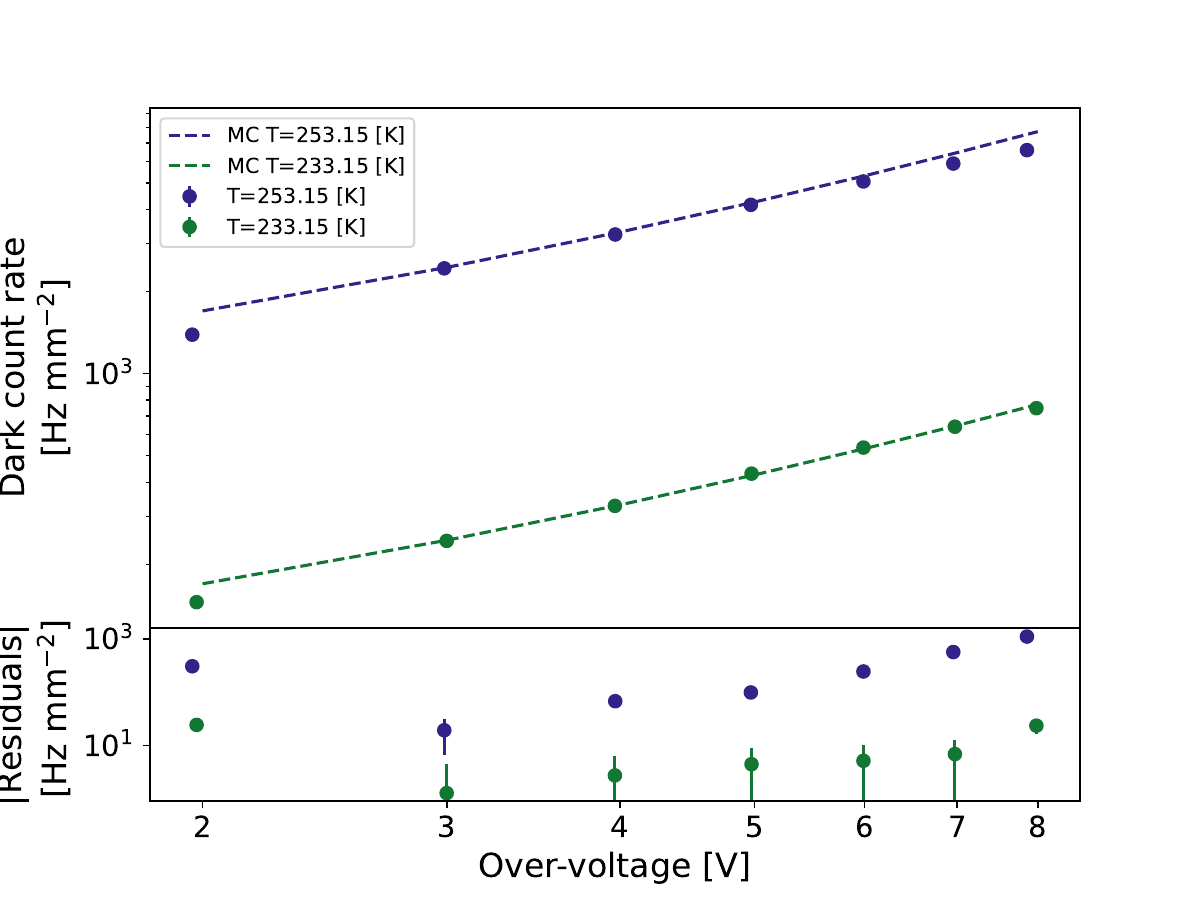}
        \caption{}
    \end{subfigure}
    \begin{subfigure}{0.45\textwidth}
        \includegraphics[width=\textwidth]{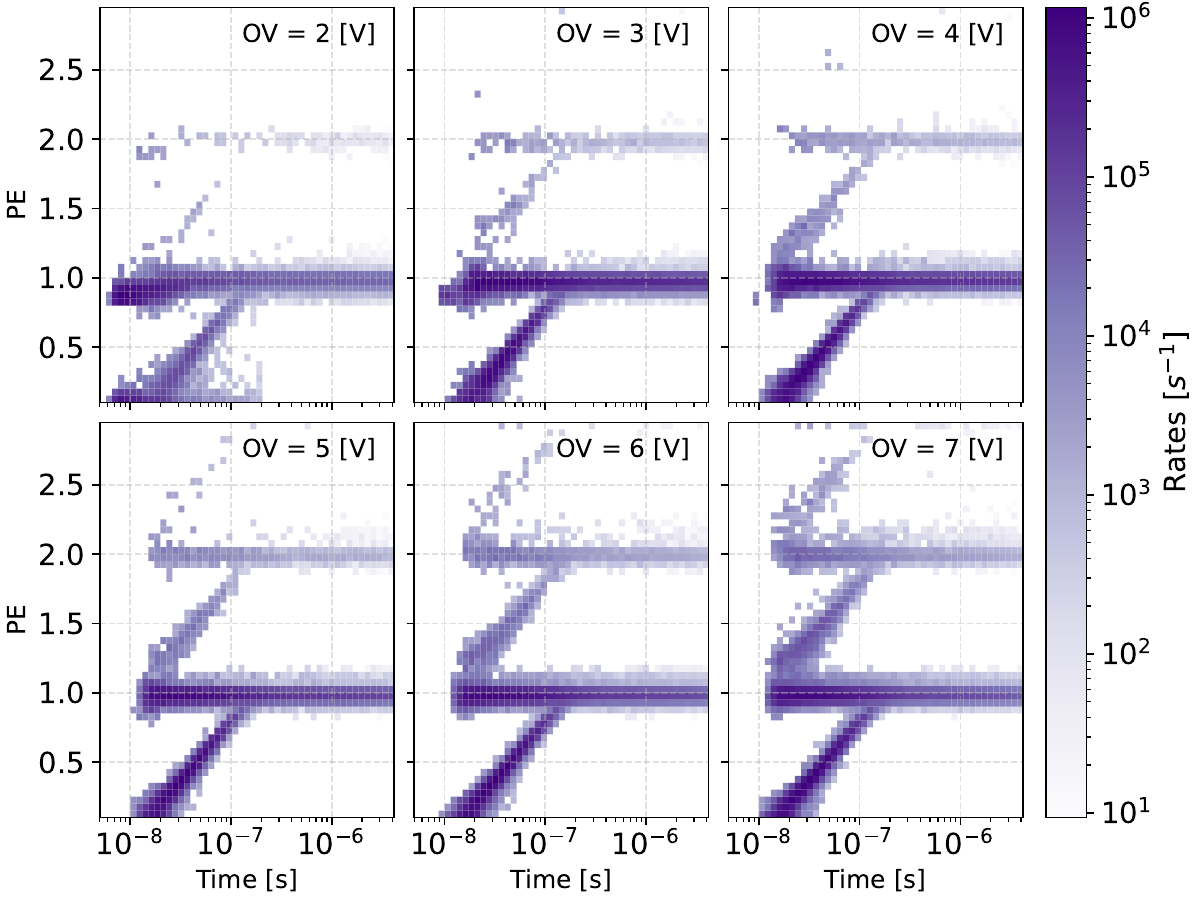}
        \caption{}
    \end{subfigure}
    \caption{(a) Estimated DCR for the MC SiPM as a function of OV for both temperatures. Because of a low signal-to-noise ratio, the 2~V DCRs are under-characterized from noise being misinterpreted as a pulse. At 253.15~K and high OVs, the bias errors are attributed to too much pile-up causing saturation. Note the x-axis is in log scale. (b) The charge vs. time with respect to the first pulse histograms at 253.15~K. The OV=2~V histogram shows strange features attributed to noise being misinterpreted as a pulse.}
    \label{fig:mc:DCR}
\end{figure}

\begin{figure}
    \centering
    \includegraphics[width=\textwidth]{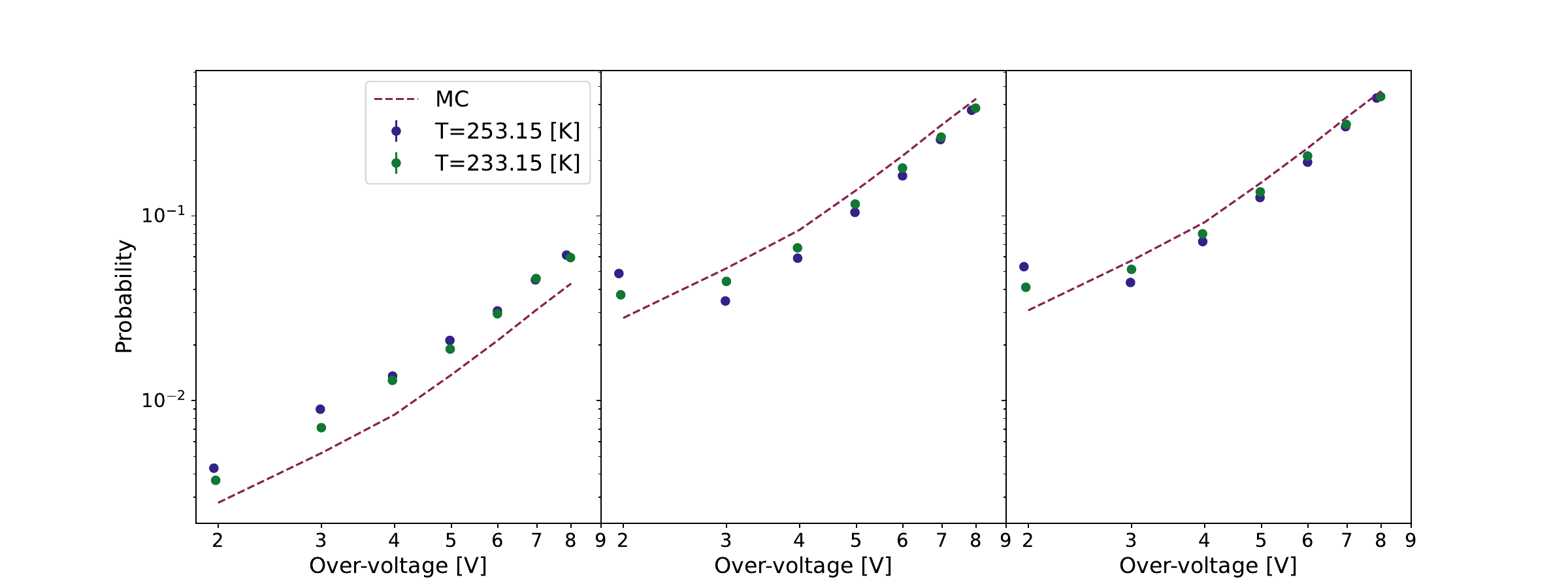}
    \caption{Estimated P$_{\text{PCA}}$ (left), P$_{\text{DCA}}$ (middle) and P$_{\text{CA}}$ (right) as a function of OV for both temperatures. P$_{\text{DCA}}$ is under-estimated because of undetected APs and low $\Delta t$ DCA pulses being mischaracterized as PCA. P$_{\text{DCA}}$ at the 2~V data point is significantly higher than expected because of the DCR error. Note the x-axis is in log scale.}
    \label{fig:mc:p}
\end{figure}

The timing parameters discussed in this section are DCR and P$_{\text{CA}}$. $\bar{\tau}_{\text{DCA}}$ is not discussed because the MC model uses several different time constants for each type of CA. DCR can be seen in left side of figure~\ref{fig:mc:DCR} while P$_{\text{PCA}}$ and P$_{\text{DCA}}$ can be seen in figure~\ref{fig:mc:p}. The chosen MC DCR values at 233.15~K are exactly one order of magnitude lower than the DCR at 253.15~K. The left side of figure~\ref{fig:mc:DCR} shows a larger difference to the expected values at 2~V which is explained by a low signal-to-noise ratio that complicates the timing extraction algorithm which is reflected in the left side of the 2~V histogram in figure~\ref{fig:mc:DCR}. The differences at high OV for 253.15~K can be explained by a failure in the unfolding algorithm caused by reduced statistics and too much pile-up. Pile-up occurs when the DCR is higher than approximately 5~kHz mm$^{-2}$. In previous attempts, the digitizer saturation (in this case simulated) heavily underestimates the calculated DCR and therefore P$_{\text{CA}}$. The origin of the mischaracterization was found to be when the CAs of the independent pulses were too numerous, which caused the CA to be indistinguishable from the original pulse up to timeframes in the few hundred ns. Nevertheless, to fully characterize the biases introduced by the analysis, a full study of DCR vs.~P$_{\text{DCA}}$ is required which is dependent on pulse shape and the probabilities of a CA affecting saturation.

All probabilities of any P$_{\text{CA}}$ were kept dependent only on OV, and P$_{\text{PCA}}$ was chosen to be exactly one order of magnitude smaller than P$_{\text{DCA}}$. The results show that P$_{\text{DCA}}$ is always underestimated. It was an expected outcome of potential AP going undetected at the wave-unfolding step from a poor signal-to-noise ratio and low $\Delta t$ DCA pulses being mischaracterized. Conversely, P$_{\text{PCA}}$ were always over-estimated compared to the MC model. P$_{\text{PCA}}$ bias errors should come from two sources, a DCR pulse pile-up with the first pulse whose probability is negligible for most DCR studied in this paper, and the mischaracterizing of DCAs.

The MC data shows that the analysis framework developed for this paper is precise enough to characterize SiPM parameters to within a few percent and the reported errors in section~\ref{sec:results} are consistently higher than the MC bias error. The circumstances where this fails can be attributed to mischaracterizations, saturation errors, or low statistics. No correction is applied to the measured values, as a more complete MC model would be required for a fair comparison. In the simplest case, a bias error of 50\% can be applied to all the P$_{\text{CA}}$ estimations to justly translate those values between experimental setups.

Improvements to the MC model are required to push knowledge of the parameters beyond the ones reported in this paper. To fully incorporate the MC data, a more accurate MC model of SiPMs must take into account a full electrical model of a SiPM pulse \cite{seifert2009}, the probability of a CA as a function of geometry \cite{rosado2015}, SPAD-to-SPAD variations, and correlation of correlations. A final consideration to further optimize the MC is the relatively slow waveform creation speed. Despite optimizations, waveforms that are a few $\upmu$s in length require a few ms of computation time. 

\section{Glossary}
\begin{itemize}
    \item sp: sample-period equal to the inverse of the acquisition frequency
    \item $V_{\text{BD}}$: Breakdown Voltage [V]
    \item OV: over-voltage, OV$ = V - V_{\text{BD}}$ [V]
    \item SPE: single photo-electron
    \item $A$: SiPM charge in arbitrary units [arb.]
    \item $Q$: SiPM charge in Coulombs [C]
    \item $\alpha$: proportional constant to turn arbitrary units in units of charge $Q = \alpha A$ [C~arb.$^{-1}$]
    \item $g_{\text{SiPM}}$: SiPM gain $g_{\text{SiPM}} = \alpha A / q_E$ [$e^-$~PE$^{-1}$]
    \item $m$: rate of change of gain vs.~over-voltage [$e^-$~PE$^{-1}$~V$^{-1}$]
    \item DCR: dark counts rate [Hz~mm$^{-2}$]
    \item CA: correlated avalanche
    \item AP: after pulse
    \item DCA: delayed CA
    \item PCA: prompt CA
    \item P$_{\text{DCA}}$: probability of a DCA
    \item P$_{\text{PCA}}$: probability of a PCA
    \item P$_{\text{CA}}$: probability of a CA. P$_{\text{CA}} =$ P$_{\text{PCA}} + $P$_{\text{DCA}}$
    \item $\bar{\tau}_{DCA}$: average arrival time with respect to the first pulse of the DCAs [ns]
\end{itemize}
\bibliographystyle{JHEP} 
\bibliography{bibliography}

\providecommand{\href}[2]{#2}\begingroup\raggedright\begin{thebibliography}{10}

\bibitem{neumeier2015}
A.~Neumeier, T.~Dandl, T.~Heindl, A.~Himpsl, L.~Oberauer, W.~Potzel et~al., \emph{Intense vacuum ultraviolet and infrared scintillation of liquid {{Ar-Xe}} mixtures}, \href{https://doi.org/10.1209/0295-5075/109/12001}{\emph{EPL} {\bfseries 109} (2015) 12001}.

\bibitem{amole2019}
{\scshape PICO} collaboration, \emph{Dark matter search results from the complete exposure of the {{PICO-60 C}}{$_{3}$}{{F}}{$_8$} bubble chamber}, \href{https://doi.org/10.1103/PhysRevD.100.022001}{\emph{Phys. Rev. D} {\bfseries 100} (2019) 022001}.

\bibitem{baxter2017}
D.~Baxter, C.J.~Chen, M.~Crisler, T.~Cwiok, C.E.~Dahl, A.~Grimsted et~al., \emph{First {{Demonstration}} of a {{Scintillating Xenon Bubble Chamber}} for {{Detecting Dark Matter}} and {{Coherent Elastic Neutrino-Nucleus Scattering}}}, \href{https://doi.org/10.1103/PhysRevLett.118.231301}{\emph{Phys. Rev. Lett.} {\bfseries 118} (2017) 231301}.

\bibitem{gundacker2020}
S.~Gundacker and A.~Heering, \emph{The silicon photomultiplier: Fundamentals and applications of a modern solid-state photon detector}, \href{https://doi.org/10.1088/1361-6560/ab7b2d}{\emph{Phys. Med. Biol.} {\bfseries 65} (2020) 17TR01}.

\bibitem{alfonso-pita2022}
{\scshape SBC} collaboration, \emph{Snowmass 2021 {{Scintillating Bubble Chambers}}: {{Liquid-noble Bubble Chambers}} for {{Dark Matter}} and {{CE}}{$\nu$}{{NS Detection}}},  \href{https://arxiv.org/abs/2207.12400}{{\ttfamily 2207.12400}}.

\bibitem{alfonso-pita2023}
{\scshape SBC} collaboration, \emph{Scintillating {{Bubble Chambers}} for {{Rare Event Searches}}}, \href{https://doi.org/10.3390/universe9080346}{\emph{Universe} {\bfseries 9} (2023) 346}.

\bibitem{hawley-herrera2024}
{\scshape SBC} collaboration, \emph{{{SBC-SNOLAB}} scintillation system and {{SiPM}} implementation for dark matter searches}, \href{https://doi.org/10.1088/1748-0221/19/01/C01023}{\emph{JINST} {\bfseries 19} (2024) C01023}.

\bibitem{acerbi2015}
F.~Acerbi, A.~Ferri, G.~Zappala, G.~Paternoster, A.~Picciotto, A.~Gola et~al., \emph{{{NUV}} silicon photomultipliers with high detection efficiency and reduced delayed correlated-noise}, \href{https://doi.org/10.1109/TNS.2015.2424676}{\emph{IEEE Trans. Nucl. Sci.} {\bfseries 62} (2015) 1318}.

\bibitem{piemonte2016}
C.~Piemonte, F.~Acerbi, A.~Ferri, A.~Gola, G.~Paternoster, V.~Regazzoni et~al., \emph{Performance of {{NUV-HD}} silicon photomultiplier technology}, \href{https://doi.org/10.1109/TED.2016.2516641}{\emph{IEEE Trans. Electron Devices} {\bfseries 63} (2016) 1111}.

\bibitem{seifert2009}
S.~Seifert, H.T.~{van Dam}, J.~Huizenga, R.~Vinke, P.~Dendooven, H.~Lohner et~al., \emph{Simulation of silicon photomultiplier signals}, \href{https://doi.org/10.1109/TNS.2009.2030728}{\emph{IEEE Trans. Nucl. Sci.} {\bfseries 56} (2009) 3726}.

\bibitem{gola2019}
A.~Gola, F.~Acerbi, M.~Capasso, M.~Marcante, A.~Mazzi, G.~Paternoster et~al., \emph{{{NUV-Sensitive Silicon Photomultiplier Technologies Developed}} at {{Fondazione Bruno Kessler}}}, \href{https://doi.org/10.3390/s19020308}{\emph{Sensors} {\bfseries 19} (2019) 308}.

\bibitem{rosado2015}
J.~Rosado, V.M.~Aranda, F.~Blanco and F.~Arqueros, \emph{Modeling crosstalk and afterpulsing in silicon photomultipliers}, \href{https://doi.org/10.1016/j.nima.2014.11.080}{\emph{NIM} {\bfseries 787} (2015) 153}.

\bibitem{du2008}
Y.~Du and F.~Reti{\`e}re, \emph{After-pulsing and cross-talk in multi-pixel photon counters}, \href{https://doi.org/10.1016/j.nima.2008.08.130}{\emph{NIM} {\bfseries 596} (2008) 396}.

\bibitem{boulay2023}
M.G.~Boulay, V.~Camillo, N.~Canci, S.~Choudhary, L.~Consiglio, A.~Flammini et~al., \emph{{{SiPM}} cross-talk in liquid argon detectors}, \href{https://doi.org/10.3389/fphy.2023.1181400}{\emph{Front. Phys.} {\bfseries 11} (2023) }.

\bibitem{acerbi2019}
F.~Acerbi, G.~Paternoster, M.~Capasso, M.~Marcante, A.~Mazzi, V.~Regazzoni et~al., \emph{Silicon {{Photomultipliers}}: {{Technology Optimizations}} for {{Ultraviolet}}, {{Visible}} and {{Near-Infrared Range}}}, \href{https://doi.org/10.3390/instruments3010015}{\emph{Instruments} {\bfseries 3} (2019) 15}.

\bibitem{aalseth2017}
C.E.~Aalseth, F.~Acerbi, P.~Agnes, I.F.M.~Albuquerque, T.~Alexander, A.~Alici et~al., \emph{Cryogenic {{Characterization}} of {{FBK RGB-HD SiPMs}}}, \href{https://doi.org/10.1088/1748-0221/12/09/P09030}{\emph{JINST} {\bfseries 12} (2017) P09030}.

\bibitem{bisogni2019}
M.G.~Bisogni, A.~Del~Guerra and N.~Belcari, \emph{Medical applications of silicon photomultipliers}, \href{https://doi.org/10.1016/j.nima.2018.10.175}{\emph{NIM} {\bfseries 926} (2019) 118}.

\bibitem{riu2012}
J.~Riu, M.~Sicard, S.~Royo and A.~Comer{\'o}n, \emph{Silicon photomultiplier detector for atmospheric lidar applications}, \href{https://doi.org/10.1364/OL.37.001229}{\emph{Opt. Lett., OL} {\bfseries 37} (2012) 1229}.

\bibitem{hoenk1992}
M.E.~Hoenk, P.J.~Grunthaner, F.J.~Grunthaner, R.W.~Terhune, M.~Fattahi and H.-F.~Tseng, \emph{Growth of a delta-doped silicon layer by molecular beam epitaxy on a charge-coupled device for reflection-limited ultraviolet quantum efficiency}, \href{https://doi.org/10.1063/1.107675}{\emph{Applied Physics Letters} {\bfseries 61} (1992) 1084}.

\bibitem{nagai2019}
A.~Nagai, C.~Alispach, A.~Barbano, V.~Coco, D.~Della~Volpe, M.~Heller et~al., \emph{Characterization of a large area silicon photomultiplier}, \href{https://doi.org/10.1016/j.nima.2019.162796}{\emph{Nuclear Instruments and Methods in Physics Research Section A: Accelerators, Spectrometers, Detectors and Associated Equipment} {\bfseries 948} (2019) 162796}.

\bibitem{dinu2017}
N.~Dinu, A.~Nagai and A.~Para, \emph{Breakdown voltage and triggering probability of {{SiPM}} from {{IV}} curves at different temperatures}, \href{https://doi.org/10.1016/j.nima.2016.05.110}{\emph{Nuclear Instruments and Methods in Physics Research Section A: Accelerators, Spectrometers, Detectors and Associated Equipment} {\bfseries 845} (2017) 64}.

\bibitem{acerbi2017}
F.~Acerbi, S.~Davini, A.~Ferri, C.~Galbiati, G.~Giovanetti, A.~Gola et~al., \emph{Cryogenic characterization of {{FBK HD}} near-{{UV}} sensitive {{SiPMs}}}, \href{https://doi.org/10.1109/TED.2016.2641586}{\emph{IEEE Trans. Electron Devices} {\bfseries 64} (2017) 521}.

\bibitem{gallina2019}
G.~Gallina, P.~Giampa, F.~Reti{\`e}re, J.~Kroeger, G.~Zhang, M.~Ward et~al., \emph{Characterization of the {{Hamamatsu VUV4 MPPCs}} for {{nEXO}}}, \href{https://doi.org/10.1016/j.nima.2019.05.096}{\emph{NIM} {\bfseries 940} (2019) 371}.

\bibitem{taylor1994}
B.N.~Taylor and C.E.~Kuyatt, ``{{NIST Technical Note}} 1297.'' https://www.nist.gov/pml/nist-technical-note-1297, 1994.

\bibitem{preston-thomas1990}
H.~{Preston-Thomas}, \emph{The {{International Temperature Scale}} of 1990 ({{ITS-90}})}, \href{https://doi.org/10.1088/0026-1394/27/1/002}{\emph{Metrologia} {\bfseries 27} (1990) 3}.

\bibitem{keithleyinstruments1998}
I.~Keithley~Instruments, \emph{Low Level Measurements: {{For}} Effective Low Current, Low Voltage, and High Impedance Measurements}, Keithley Instruments, 5~ed. (1998).

\bibitem{pedregosa2011}
F.~Pedregosa, G.~Varoquaux, A.~Gramfort, V.~Michel, B.~Thirion, O.~Grisel et~al., \emph{Scikit-learn: {{Machine}} learning in {{Python}}}, {\emph{J. Mach. Learn. Res.} {\bfseries 12} (2011) 2825}.

\bibitem{xu2022}
D.C.~Xu, B.D.~Xu, E.J.~Bao, Y.Y.~Wu, A.Q.~Zhang, Y.Y.~Wang et~al., \emph{Towards the ultimate {{PMT}} waveform analysis for neutrino and dark matter experiments}, \href{https://doi.org/10.1088/1748-0221/17/06/P06040}{\emph{JINST} {\bfseries 17} (2022) P06040}.

\bibitem{peterson2021}
J.H.~Peterson, \emph{Developments in waveform unfolding of {{PMT}} signals in future {{IceCube}} extensions}, \href{https://doi.org/10.1088/1748-0221/16/09/C09032}{\emph{JINST} {\bfseries 16} (2021) C09032}.

\bibitem{lgallego2013}
{L Gallego}, {J Rosado}, {F Blanco} and {F Arqueros}, \emph{Modeling crosstalk in silicon photomultipliers}, \href{https://doi.org/10.1088/1748-0221/8/05/P05010}{\emph{JINST} {\bfseries 8} (2013) P05010}.

\bibitem{butcher2017}
A.~Butcher, L.~Doria, J.~Monroe, F.~Reti{\`e}re, B.~Smith and J.~Walding, \emph{A method for characterizing after-pulsing and dark noise of {{PMTs}} and {{SiPMs}}}, \href{https://doi.org/10.1016/j.nima.2017.08.035}{\emph{NIM} {\bfseries 875} (2017) 87}.

\bibitem{serra2011a}
N.~Serra, G.~Giacomini, A.~Piazza, C.~Piemonte, A.~Tarolli and N.~Zorzi, \emph{Experimental and {{TCAD}} study of breakdown voltage temperature behavior in {{n$^+$/p}} {{SiPMs}}}, \href{https://doi.org/10.1109/TNS.2011.2123919}{\emph{IEEE Trans. Nucl. Sci.} {\bfseries 58} (2011) 1233}.

\bibitem{liu2022}
M.~Liu, B.~Liu, J.~Hu, D.~Li, J.~Ma, Z.~Chu et~al., \emph{A 16-{{Channel}} analog {{CMOS SiPM}} with on-chip front-end for {{D-ToF LiDAR}}}, \href{https://doi.org/10.1109/TCSI.2022.3160265}{\emph{IEEE Trans. Circuits Syst. Regul. Pap.} {\bfseries 69} (2022) 2376}.

\bibitem{ieki2019}
K.~Ieki, T.~Iwamoto, D.~Kaneko, S.~Kobayashi, N.~Matsuzawa, T.~Mori et~al., \emph{Large-area {{MPPC}} with enhanced {{VUV}} sensitivity for liquid xenon scintillation detector}, \href{https://doi.org/10.1016/j.nima.2019.02.010}{\emph{Nuclear Instruments and Methods in Physics Research Section A: Accelerators, Spectrometers, Detectors and Associated Equipment} {\bfseries 925} (2019) 148}.

\bibitem{garutti2021}
E.~Garutti, R.~Klanner, J.~Rolph and J.~Schwandt, \emph{Simulation of the response of {{SiPMs}}; {{Part~I}}: {{Without}} saturation effects}, \href{https://doi.org/10.1016/j.nima.2021.165853}{\emph{NIM} {\bfseries 1019} (2021) 165853}.

\end{thebibliography}\endgroup

\end{document}